\documentclass[11pt,a4paper]{article} 

\usepackage{jheppub,esint,shuffle,psfrag}
\usepackage{amsmath, amssymb,amsthm}
\usepackage{amsthm}
\usepackage{bm,environ,mathrsfs,array,arydshln}
 \usepackage{tensor} 						

\newcommand{\bs}{\begin{shaded}}
\newcommand{\es}{\end{shaded}}
\def\ba#1\ea{\begin{align}#1\end{align}}		
\newcommand{\be}{\begin{equation}}
\newcommand{\ee}{\end{equation}}
\newcommand{\mc}{\mathcal }

\newcommand{\la}{\label}
\newcommand{\eps}{\varepsilon}

\DeclareMathOperator{\tr}{\text{tr}}

\newcommand{\cf}{\textit{cf.} }
\newcommand{\ie}{\textit{i.e.} }

\makeatletter
\DeclareFontFamily{OMX}{MnSymbolE}{}
\DeclareSymbolFont{MnLargeSymbols}{OMX}{MnSymbolE}{m}{n}
\SetSymbolFont{MnLargeSymbols}{bold}{OMX}{MnSymbolE}{b}{n}
\DeclareFontShape{OMX}{MnSymbolE}{m}{n}{
<-6>  MnSymbolE5
   <6-7>  MnSymbolE6
   <7-8>  MnSymbolE7
   <8-9>  MnSymbolE8
   <9-10> MnSymbolE9
  <10-12> MnSymbolE10
  <12->   MnSymbolE12
}{}
\DeclareFontShape{OMX}{MnSymbolE}{b}{n}{
<-6>  MnSymbolE-Bold5
   <6-7>  MnSymbolE-Bold6
   <7-8>  MnSymbolE-Bold7
   <8-9>  MnSymbolE-Bold8
   <9-10> MnSymbolE-Bold9
  <10-12> MnSymbolE-Bold10
  <12->   MnSymbolE-Bold12
}{}

\let\llangle\@undefined
\let\rrangle\@undefined
\DeclareMathDelimiter{\llangle}{\mathopen}%
 {MnLargeSymbols}{'164}{MnLargeSymbols}{'164}
\DeclareMathDelimiter{\rrangle}{\mathclose}%
 {MnLargeSymbols}{'171}{MnLargeSymbols}{'171}
\makeatother

\newmuskip\pFqmuskip

\newcommand*\pFq[6][8]{%
  \begingroup 
  \pFqmuskip=#1mu\relax
  \mathchardef\normalcomma=\mathcode`,
  \mathcode`\,=\string"8000
  \begingroup\lccode`\~=`\,
  \lowercase{\endgroup\let~}\pFqcomma
  {}_{#2}F_{#3}{\left[\genfrac..{0pt}{}{#4}{#5};#6\right]}%
  \endgroup
}
\newcommand{\pFqcomma}{{\normalcomma}\mskip\pFqmuskip}

\def\XXint#1#2#3{{\setbox0=\hbox{$#1{#2#3}{\int}$}
     \vcenter{\hbox{$#2#3$}}\kern-.5\wd0}}

\def \rmP {{\rm p}}   \def \rmNP {{\rm np}}

\newcommand\re[1]{(\ref{#1})}
\def \GG {{\cal F}}

\newcommand\lr[1]{{\left({#1}\right)}}

\newcommand{\sql}{\sqrt\lambda}
\newcommand{\gs}{g_{\text{s}}}

\newcommand{\vev}[1]{\langle  #1 \rangle}

\newcommand{\gym}{g_{_{\small {\rm YM}}}}

\def \no { \nonumber}
  
\def \foot{\footnote}\def \ci{cite}\def \l {\lambda}\def \iffa {\iffalse}
\def \RR {{R}} \def \ov {\over }\def \a  {\alpha} \def \ha {{1\ov 2}}
\def \ed {\end{document}}

\def \l {\lambda}
\def\foot{\footnote}
\def \adss {${\rm AdS}_5 \times S^5~$ }
\def \ov {\over}
\def \tr {{\rm tr\,}}
\def \ha {{1 \over 2}}

\def \ci{\cite}

\def \N  {{\cal N}}
\def \aa  {{\rm a}}
\def \te {\textstyle} 
\def \Z {{\cal Z}}
\def \RR {{L}}
\def \aa {{\rm a}}   
\def \eps {\epsilon}\def \rr {{\rm r}}
\newcommand{\rf}[1]{(\ref{#1})}
\def \Z  {\mathbb{Z}}

\def \four {\frac{1}{4}}

\def \ori {{\rm ori}}

\def \N {{\cal N}}

\def \ori {{\rm ori}}

\def \N {{\cal N}}

\def \Z  {\mathbb{Z}}\def \l {\lambda}

  \def \FA {FA-orientifold\ } 

 \def \half {\tfrac{1}{2}}
  \def \Z {{\cal Z}}

 \def \rr  {{\rm r}}

\newcommand{\branching}[3]{(\bm{#1}, \bm{#2})_{#3}}
\def \Vol {{\rm Vol}}

\def \del {\partial}
\def \L  {\Lambda}
\def \four {\tfrac{1}{4}}
\def \K {{\rm k}}
\def \LL  {{\cal L}}
\def \Z {{\mathbb Z}}
\def \OO {{\cal O}}
\def \aa{{\rm a}}  \def \cc {{\rm c}}
\def \ZZ {{\mathbb Z}}
\def \ed {\newpage \bibliography{BKT-Biblio} \bibliographystyle{JHEP} \end{document}}

\def \betat {{b}}

\def \FF {{\rm F}}

\begin{document}

\vspace{-4cm }

\begin{flushleft}
 \hfill \parbox[c]{50mm}{
 IPhT--T22/079
 \\
 Imperial-TP-AT-2022-05}
\end{flushleft}
\author{M. Beccaria$^a$, G.P. Korchemsky$^{b,c}$ and A.A. Tseytlin$^{d,}$\footnote{Also on leave from  Inst. for Theoretical and Mathematical Physics (ITMP)  and Lebedev Inst.}}
\affiliation{
$\null$
$^a$Universit\`a del Salento, Dipartimento di Matematica e Fisica \textit{Ennio De Giorgi},\\ 
\phantom{a} and I.N.F.N. - sezione di Lecce, Via Arnesano, I-73100 Lecce, Italy
\\		
$\null$
$^b${Institut de Physique Th\'eorique\footnote{Unit\'e Mixte de Recherche 3681 du CNRS}, Universit\'e Paris Saclay, CNRS,\\ \phantom{a}  91191 Gif-sur-Yvette, France}  
\\
$\null$
$^c${Institut des Hautes \'Etudes Scientifiques, 91440 Bures-sur-Yvette, France}  
\\
$\null$
$^d${Blackett Laboratory, Imperial College London,  SW7 2AZ, U.K.}
}

\title{Exact  strong coupling  results  in   $\cal N$=2  \\  $Sp(2N)$ superconformal gauge theory\\
  from localization
}

\abstract{
We apply the localization technique to compute the free energy on  four-sphere and the  
circular BPS Wilson loop in the four-dimensional $\cal N$=2 superconformal    $Sp(2N)$ gauge  theory 
 containing vector multiplet coupled to  four hypermultiplets in  fundamental representation and one hypermultiplet in rank-2  antisymmetric   representation.  This theory is unique among  similar  $\cal N$=2  superconformal models that are planar-equivalent to $\cal N$=4  SYM in that the corresponding localization matrix model  has  the interaction   potential containing single-trace terms only.  We exploit this property to show that, to any order in large $N$ expansion and  an arbitrary 't Hooft coupling $\lambda$, the free energy and the Wilson loop satisfy Toda lattice equations. Solving these  equations at strong coupling, we  find   remarkably simple expressions for these observables which include all corrections in $1/N$ and $1/\sqrt{\lambda}$. We also compute the leading exponentially suppressed ${\cal O}(e^{-\sqrt{\lambda}})$ corrections and consider a generalization  to the case when  the fundamental  hypermultiplets  have  a  non-zero mass. The  string theory dual of this $\cal N$=2  gauge  theory  should  be a particular orientifold of  AdS$_5 \times S^5$   type IIB string  and we  discuss  the  string theory interpretation of  the obtained strong-coupling results.}

\maketitle
\flushbottom
\setcounter{footnote} 0

\section{Introduction  and summary}

It was recently appreciated that localization \cite{Pestun:2007rz,Pestun:2016zxk} provides an important tool to study AdS/CFT duality beyond the planar limit in 
a class of simplest $\N = 2$ superconformal models that are planar-equivalent  to $\N=4$ SYM theory  
(see, in particular, \cite{Beccaria:2020hgy,Beccaria:2021ksw,Beccaria:2021vuc,Beccaria:2021ism,Beccaria:2022ypy,Bobev:2022grf}).\foot{See also \ci{Fiol:2014fla,Fiol:2015mrp,Fiol:2020bhf}
  and \cite{Billo:2021rdb,Billo:2022xas,Billo:2022fnb}     for related computations of  special correlators of BPS operators.} 
Using localization  matrix model representation  for  some  special observables like free energy on four-sphere  or circular BPS  Wilson loop 
one can  find    their $1/N$, large 't Hooft coupling $\l$ expansions  and interpret the resulting series as perturbative expansion 
in terms of  the dual string theory parameters -- string coupling $\gs$ and string tension $T$
\be
\la{1}
\gs =  \frac{\lambda}{4\pi N},\qquad\qquad  \quad T = \frac{\RR^2}{2\pi \a'}=\frac{\sqrt\lambda}{2\pi}\ , \ \ \ \ \ \ \qquad \l=\gym^2 N \ .   
\ee
The  conceptually  
 simplest example is provided  by the $\ZZ_2$ orbifold  of the $SU(2N)$ $\N=4$  SYM theory (i.e.  the  $SU(N) \times SU(N)$  $\N=2$ gauge theory with bi-fundamental hypermultiplets and  equal couplings)  which is 
 dual to string theory on  AdS$_5 \times S^5/\ZZ_2$. 
 However, the corresponding  localization 2-matrix model is rather complicated (with the  interaction potential
   containing double-trace terms). As a result,   the strong-coupling  expansion of only the leading $1/N^2$ term in the free energy and the Wilson loop  expectation value  was  so far   worked out explicitly 
\ci{Beccaria:2021ksw,Beccaria:2022ypy}.

Here we will focus on what turns out  to be the simplest representative 
in the  family  of   similar   $\N=2$  superconformal 
models  that are  planar-equivalent to $\N=4$ super Yang-Mills theory: 
  the  $\N=2$ $Sp(2N)$ gauge theory coupled to  four  hypermultiplets  in the  fundamental 
representation 
 and  one  hypermultiplet  is the rank 2 antisymmetric  representation  of $Sp(2N)$.

This gauge  theory can be  ``engineered''   on a collection 
of $2N$ D3-branes, 8 D7-branes and one O7-plane   \ci{Sen:1996vd,Dasgupta:1996ij, 
Banks:1996nj,Aharony:1996en,
Douglas:1996js}.\foot{
The 4 fundamental hypers  are massless modes of 
 strings stretched between the D3- and D7-branes. O7 plane is required for stability (conformal invariance)   bringing in 
 the antisymmetric hyper that 
 arises from the action of the orientifold projection on the fields  corresponding to directions  transverse to the D3-branes but parallel to the D7-branes.
  The  Coulomb branch  corresponds to  giving   an expectation value 
  to the  complex scalar of $ \N = 2$ vector multiplet related to separation 
  of  D3-branes  from D7-branes  and  the fixed plane. 
  One  Higgs branch is  parametrized by  expectation values of the 
  scalars of the  antisymmetric hyper (representing motion of  D3-branes 
  in the transverse directions within  D7-branes). The  second Higgs branch  is parameterized by the fundamental scalars and corresponds to dissolving  D3-branes inside D7-branes  and may  be  described in terms of   gauge instanton moduli space.
}
  The   corresponding dual  string theory  is then expected to be  
 a special  orientifold of   type IIB superstring   on  AdS$_5 \times S^5$ \ci{Fayyazuddin:1998fb,Aharony:1998xz,Park:1998zh,Blau:1999vz,Ennes:2000fu}.
The dual description  does not   explicitly involve  D7-branes, but due to orientifolding 
there is  also an open-string sector  in addition to a closed-string  one  (like  in   type I  theory).

This   theory  (referred to as the  ``FA-orientifold''   model  in \ci{Beccaria:2021ism}) 
 is unique in that the  interaction potential  in the localization matrix model 
 contains only single-trace terms. 
This  leads   to    substantial technical simplifications compared to other similar $\N=2$ 
superconformal models with  
localization matrix model  having double-trace potentials. 
As a result, the large $N$ expansion of the free energy $F^{\N=2}_N$ and  the 
expectation value of the circular BPS Wilson loop $  W^{\N=2}_N$
can be worked out rather explicitly    for  an arbitrary 't Hooft coupling \ci{Beccaria:2021ism}.

In virtue of the planar equivalence, the large $N$ expansion of both quantities is 
\ba\notag
\la{18} 
& F_N^{\N=2} = F_N^{\N=4} + N\, \FF_1(\l) + \FF_2(\l) + {1\over N} \FF_3(\l) + \OO\big({1\ov N^2}\big)\,,
\\ 
& W^{\N=2}_N =   W^{\N=4}_N  +\Delta  W^{(1)}(\l)  +\frac{1}{N}\Delta  W^{(2)} (\l)  +  \OO \big({1\ov N^2}\big) \ , 
\ea
where the leading term is   the  $\mathcal N=4$   $Sp(2N)$ SYM  result and 
the subleading   ones  
are suppressed by powers of $1/N$.
The free energy and the Wilson loop in $\mathcal N=4$ $Sp(2N)$ theory are given by the well-known expressions  \ci{Fiol:2014fla,Giombi:2020kvo}
\ba \la{4}\notag
& F_N ^{\N=4} =   -   N ( N +\half)  \log \l\    + C_N^{\N=4} \ , 
\\
& W^{\N=4}_N  
= 2\, e^{\frac{\lambda}{16N}} \sum_{n=0}^{N-1}L_{2n+1}\big(-\tfrac{\lambda}{8N}\big)\,,
\ea
where  the $N$-dependent  
constant $C_N^{\N=4}$ can be expressed in terms of  Barnes $G-$function (see \rf{Z4} below)
and $L_{2n+1}(x)$ is the  Laguerre polynomial.

\subsection*{Toda lattice  equation}   

Using the localization matrix model representation for  the free energy $F_N^{\N=2}$, 
one  can show  \ci{Beccaria:2021ism}  that the leading non-planar correction 
$\FF_1(\lambda)$ in \rf{18} admits a compact integral  representation in terms  of  Bessel function (see \rf{F0} below). 
As was found  in \ci{Beccaria:2021ism}, 
 the subleading corrections in \rf{18} have an interesting iterative structure. 
Namely,   
the functions $\FF_k(\l)$ (with $k\ge 2$) in the free energy can be expressed in terms of the leading function $\FF_1(\lambda)$ as
\ba\notag
& 
 \FF'_{2}  = \tfrac{1}{4}(\lambda \FF_1)''  -\tfrac{1}{4}\l\big[(\lambda\, \FF_1)''\big]^{2} \,,
 \la{8}
 \\[2mm] 
& \FF_{3}  =   \tfrac{1}{48}\l^2\big( \l \FF_1\big)'''  - \tfrac{1}{16}\l^2\big[ \big(\l \FF_1\big)''\big]^{2}  +\tfrac{1}{24}\l^3\big[ \big(\l \FF_1\big)''\big]^{3}\ , \qquad \ \dots\ , 
 \ea
 where prime denotes a derivative with respect to $\lambda$.
For the circular Wilson loop in \rf{18}, the functions $\Delta W^{(k)}$ (with $k\ge 1$) satisfy similar relations
\ba\notag
\la{22} 
& (\Delta W^{(1)})'=  -\tfrac{1}{8}\,\l\,   W^{(0)}(\l) \big(\lambda \FF_1\big)''\ , 
\\[2mm]  
& \Delta W^{(2)} = -\tfrac{1}{32}  \lambda^{2}\,  W^{(0)}(\l) \Big[    \big(\lambda \FF_1\big)'' -  \l \big((\lambda \FF_1)''\big)^{2} \Big]
\ , \qquad  \dots\ , 
\ea
where  $W^{(0)}   =   \frac{4}{\sql} I_{1}(\sql)$ is the leading term in the large $N$ expansion of the Wilson loop in $\mathcal N=4$ theory,
$ W^{\N=4}_N=N W^{(0)} + O(N^0)$.  

The relations \rf{8} and \rf{22} were derived in \ci{Beccaria:2021ism} by examining the 
large $N$ expansions of the free energy and the Wilson loop 
in $\mathcal N=2$ $Sp(2N)$ theory at weak coupling. They are expected to hold for an arbitrary 't Hooft coupling. Being supplemented with the expression for $\FF_1(\lambda)$, they allow one  to compute subleading corrections in \rf{18} for any $\lambda$. 

In this paper we explain the origin of the relations \rf{8} and \rf{22}. 
We exploit the fact that the  localization $Sp(2N)$ matrix model representation  here   contains  only  a single-trace  interaction potential 
 to show that the free energy and the Wilson loop satisfy  discrete Toda-like equations\foot{Though not directly related, let us note that 
 the strategy of  using Toda-like recursions to control the full $1/N$ expansion was previously applied in \cite{Mangazeev:2010vu}
to compute the form factor expansions in the 2d Ising model.}
\begin{align} \notag\label{Toda-int}
& \partial^2_y F_N = -\exp\left( -F_{N+1}+2 F_N -F_{N-1}\right)\,,
\\[2mm]
& \partial_{y}^2 \, W_N  = - \left( W_{N+1}-2 W_{N}+ W_{N-1} \right)\partial_{y}^2F_N\,,
\end{align}
where 
\be \la{yla}
y\equiv {(4\pi)^2\ov g^2_{_{\rm YM}}} \ , \qquad \qquad \ \ \   \lambda=(4\pi)^2{N\ov y} \ . \ee
The  relations \rf{Toda-int}   are not sensitive to a detailed 
form of the  single-trace  interaction potential  in the localization matrix model representation for  $F_N$ and $W_N$ and, as a consequence, they hold both in the  $\mathcal N=4$ and $\mathcal N=2$  $Sp(2N)$ theories.
 The expressions on the right-hand side of \rf{Toda-int} involve functions defined for the same $y$ and $N$ shifted by $ \pm 1$. 
 In terms of the  't Hooft coupling constant, this corresponds to replacing $\lambda$ with ${N\pm 1\ov N}\lambda$.

It is straightforward to verify  that the relations \rf{Toda-int} are indeed satisfied  in $\mathcal N=4$ theory (cf.  \rf{4}).
 In the $\mathcal N=2$ theory, 
we  reproduce \rf{8} and \rf{22} by replacing $F_N\equiv F_N^{\N=2}$ and $W_N\equiv W_N^{\N=2}$ with their $1/N$ 
expansions  \rf{18} and  comparing the $1/N^k$ coefficients  on both sides of \rf{Toda-int}.

\subsection*{Strong coupling expansion}  

Solving the Toda equations \rf{Toda-int} at strong coupling (i.e. in   large $\l$ expansion), 
we find that the free energy $F_N^{\N=2}(\l)$ 
can be naturally separated into  
 ``perturbative'' and ``nonperturbative'' contributions  
\be \la{9}
F_N^{\N=2}(\l) = F_{\rmP}(N,\l) + F_{\rmNP}(N,\l)  \ . 
\ee
Here   each term in  the large $N$ expansion of $F_{\rmP}$ is
 given by a series in $1\ov \sqrt\l$ (which should  correspond to 
 $\a'$-corrections in dual  string theory). In contrast, the $O(1/N^k)$ terms  in 
$F_{\rmNP}$  are given by  sums of  
 powers of exponentially small factors $\OO(e^{-\sql})$ with coefficients depending on $1\ov \sqrt\l$.  
  
 The presence of  the latter  exponential corrections 
is reflecting  an  asymptotic (but,  it turns out,   Borel-summable) nature of the strong coupling expansion. 
On the dual  string-theory side these  corrections 
 should  correspond to  ``world-sheet instanton'' contributions that may 
 be  non-trivial due to the presence  of  a non-contractible 2-cycle in the $\ZZ_2$ 
 orbifold of $S^5$ which is part of the orientifold projection.

Remarkably, the perturbative  contribution  in  the free energy \rf{9} can be found in a closed  form  \ci{Beccaria:2021ism} (see section \ref{S2})  
\begin{align}\label{11}
F_{\rmP}(N,\l) & = (N+\tfrac{3}{4})(N+\tfrac{1}{4})\,\log \Big( {1\over \l}   +  {\log 2 \over  2\pi^2 N} \Big) 
 +C_N^{\N=2}
-\frac{\pi^{2}}{2}\frac{N}{\l}\,,
\end{align}
where the constant $C_N^{\N=2}$ is given in  \rf{resum1} below. 
The logarithmic term in \rf{11} sums up an infinite series of corrections in $1/N$ proportional to powers of  $\log 2$.  
The relation \rf{11} 
 suggests to redefine the strong-coupling  expansion parameter  $\l$ as 
\begin{align}\label{12}
{1\over \l'} \equiv  {1\over \l} + {\betat\over N } \ , \ \ \ \ \ \ \  \qquad   \betat = \frac{\log 2}{2\pi^{2}}   \,.
\end{align}
Similar redefinitions of gauge coupling previously appeared in the strong-coupling calculation   
of cusp anomalous dimension \cite{Basso:2007wd} and
 octagon correlator \cite{Belitsky:2020qrm,Belitsky:2020qir} in $\N=4$ SYM theory.

Setting  $\l'=  \gym'^{2} N$  this  redefinition can be interpreted as originating from  a 
finite  gauge 
coupling  renormalization\foot{Similar 1-loop finite renormalizations of couplings  appear  in  some supersymmetric theories:  in prepotential calculation at finite $N$ in $SU(N)$ $\N=2$ theories with flavor \cite{Dorey:1996bn,DHoker:1999yni}  and also in  some superconformal models \cite{Billo:2010mg,Billo:2013fi} (cf. also \ci{Shifman:1998zy}).  } 
\be 
 \gym'^{2} =  {\gym^{2} \over 1 +  \betat \, \gym^{2}} \ . \la{122} 
\ee
In the  localization matrix model description, the redefinition \rf{12} is closely related to the 
asymptotic behaviour of the interaction potential   at large $X$ (see \rf{11.2}  below), namely, $S_{\rm int}(X)\to b\, \tr X^2$.
 This  
produces an extra contribution to the coefficient of the   Gaussian term ${1\ov \gym^2} \tr X^2$ in the matrix model action
 leading to \rf{122}. 
 
 Rewriting $F_{\rmP}(N,\l)$ in terms of $\l'$ we find that the leading $\OO(N^2\log\l')$ term in \rf{11}  is the same as in $\mathcal N=4$ theory, Eq.~\rf{4}. 
Surprisingly,  viewed as a function of $N$ and $\l'$, the perturbative part  of the $\N=2$   free energy \rf{11} 
thus receives only  $\OO(N)$ and $\OO(N^0)$   but no higher-order  $\OO(1/N)$ corrections!
 
At large $N$ 
the leading  term  in the  nonperturbative correction to the free energy \rf{9} scales as $F_{\rmNP} \sim N \l^{-1/4} e^{-\sqrt{\l}}$. 
{ The Toda equation \re{Toda-int} allows us to compute systematically subleading nonplanar corrections to $ F_{\rmNP} $. We find that for sufficiently large $\lambda'$ these corrections  are dominated by terms of the form $\OO(({{\lambda'}^{3/2}/  N^2})^k) $. Such terms can be summed to all orders in $k$ to give  the  following resummed expression }
\begin{align} \label{Fnp-int}
F_{\rmNP}(N,\l)   
&= \tfrac{8\sqrt 2}{\pi^{3/2}}\,N    \,{\l'}^{-1/4} \, e^{-\sqrt{\l'}-{{\lambda'}^{3/2}\over 384 N^2}}  +\dots\,,
\end{align}
where dots denote  contributions   of subleading corrections.\foot{Here the   iteration of the  leading $1/N$ (or ``open-string loop'') correction  is resummed   by shifting $\l \to \l'$
like in \rf{11} 
 while  the resummation of the 
iteration of the  leading $1/N^2$ (or ``closed string loop'') correction    is represented by the factor $e^{-{{\lambda'}^{3/2}\over 384 N^2}}$.   
}
   
The Toda lattice equations \rf{Toda-int} can be effectively applied to derive the strong coupling expansion of the 
expectation value of the circular BPS Wilson loop. 
In the $\mathcal N=4$  SYM theory one finds from \rf{4}  that 
summing up the  leading large $\l$ terms at each order in $1/N$ gives \ci{Beccaria:2021ism} \foot{\la{foot4}The   factor  $\exp[{{\lambda^{3/2}\over 96 (2N)^2}}] = \exp ( \frac{\pi}{48} \frac{ \gs^2}{T})$  which is the same as in the $SU(2N)$ case \ci{Drukker:2000rr,Giombi:2020mhz}  
may be given  a string theory interpretation as a sum of  separated  one-handle  contributions to the disc partition function.}
\begin{align}\label{W-N4-int}
W_N^{\mathcal N=4} (\l)= N\sqrt{\tfrac{8}{ \pi}} \, \lambda^{-3/4} \, (1 + \tfrac{ \sql}{8 N}) \, e^{\sqrt\lambda+{\lambda^{3/2}\over 384N^2}+\dots}  +\OO\Big(e^{-\sqrt\lambda -{\lambda^{3/2}\over 384N^2}}\Big)\,,
\end{align}
where the second (nonperturbative) term is suppressed relative 
 to the first (perturbative) term by  an exponentially small factor $e^{-2\sqrt\l}$. 
Here dots stand again for subleading at large $\l$ contributions.\foot{\la{foot7}The term 
 ${\lambda^{3/2}\over 384N^2}$ in exponent of \rf{W-N4-int} comes from resummation of $\OO\big((\lambda^{3/2}/N^2)^k\big)$ corrections to all orders in $k$. Using  the exact expression \rf{4}
for $W_N^{\mathcal N=4} (\l)$ (valid for any $N$ and $\l$) one can verify that  
this
 gives a good approximation to the exact  result  in the formal double-scaling limit 
 $N\to \infty, \ \l\to \infty$  with $ {\lambda^{3/2}/N^2}=\text{fixed}$. }
 In
 the  $\mathcal N=2$ theory, the Wilson loop takes the form similar to \rf{W-N4-str}
\be\label{W-int} 
W_N^{\N=2}(\l)  = W_{\rmP}(N,\l) + W_{\rmNP}(N,\l)  \, ,
\ee
where the second term is exponentially small at strong coupling compared to the first one.
We show below that, up to  redefinition  of the coupling \rf{12}  and  an   extra rescaling  $\l'\to \l'  {N+ {1\ov 2} \ov N}$      combined with the  shift $N\to N+\tfrac12$, the perturbative term $W_{\rmP}(N,\l)$ coincides with the perturbative part of  the 
$\mathcal N=4$  theory result \rf{W-N4-str}
\begin{align}\label{Wp-int}
W_{\rmP}(N,\lambda) &=W^{\mathcal N=4}_{N+\frac12}\Big(\lambda'\frac{N+\frac12}{N}\Big) \,.
\end{align}
Recalling the definition \rf{12} and replacing $\lambda'=\gym^2 N/(1+b \gym^2 )$, the expression on the right hand side of \re{Wp-int} can be obtained from
 $W^{\mathcal N=4}_{N}\big(\lambda=\gym^2 N\big)$  simply by  
 replacing $N\to N+\tfrac12$ with $g_{_{\rm YM}}^2$ left intact.

In  
contrast  to the $\mathcal N=4$ expression \rf{W-N4-int}, the nonperturbative correction in  \rf{W-int} is  suppressed only 
by the factor  $e^{-\sqrt\l}$ as compared to $W_{\rmP}(N,\lambda)$.  
Explicitly, it is given by (see section \ref{S3})
\begin{align}\label{Wnp-int}
W_{\rmNP}(N,\lambda) = - \frac{1}{ \pi^2}\l'  + \OO(\sqrt{\lambda'})\,,
\end{align}
where  all $1/N$ corrections are  again   absorbed into the coupling $\l'$ defined in \rf{12}.

\subsection*{Massive deformation} 
 
One interesting generalization of the $\N=2$ $Sp(2N)$ model that we  consider below is to introduce a  mass for the  fundamental hypermultiplets
(cf. \ci{Aharony:1996en,Douglas:1996js}).\foot{On the  string side  this  should  correspond 
 to  introducing a separation between D3 and D7+O7 branes in 
 the directions transverse to the D7-branes (cf.  a related setup  discussed in \ci{Karch:2015kfa}).}
The dependence on this mass  parameter  is straightforward to include in the localization matrix model
potential   \ci{Pestun:2007rz}.
As we   find  below, the free energy $F_N^{\N=2}$ and the Wilson loop $W_N^{\N=2}$ have a 
 remarkably simple dependence on the mass parameter  $m$ at strong coupling. 

In particular,   the perturbative part of the free energy \rf{11}  becomes (see \rf{resum1})\foot{Note that 
the last  $\frac{N}{\l}= {1\ov \gym^2}$   term here  or  in \rf{11}   may be  written also  as $\frac{N}{\l'}$  
at the same time    redefining   the constant $C_N^{\N=4}$  by a $ \log 2$ term  (cf. \rf{122}, \rf{resum1}).}  
\ba 
F_{\rmP}(N,\l,m) = & - (N+\tfrac{3}{4}+2m^{2})(N+\tfrac{1}{4}+2m^{2})\log {\l' } +  C_N^{\N=2}(m^2)\no \\
& \qquad -  \left[ 1 + \tfrac{32}{ 3} m^2 (1+ m^2)\right] \tfrac{\pi^{2}}{2}  \tfrac{N}{\l}\,. \la{16} 
\ea
 Notice that the coefficient of the logarithmic term in \rf{16} can be obtained from the one  in \rf{11}
  by  the shift $N \to N + 2 m^2$. 

The dependence of the perturbative part of the  Wilson loop expectation value 
on the mass $m$ can  be obtained from \rf{Wp-int}  also  by   the same shift
  $N \to N + 2 m^2$  (with $\gym'$  fixed)
\begin{align}
W_{\rmP}(N,\lambda,m) &=W^{\mathcal N=4}_{N+\frac12+2m^2}\Big(\lambda'\frac{N+\frac12+2m^2}{N}\Big) \,.
\end{align}
In the localization matrix model description, the shift $N \to N + 2 m^2$ naturally follows from the structure of 
 $m$-dependence of the  interaction potential (see  section \ref{S2-4} below). 
 The meaning of this  shift  on the string theory side is an open question. 

For the nonperturbative part of $F_N^{\N=2}$ and $W_N^{\N=2}$, the 
generalization to nonzero $m$ amounts to inserting the factor of $\cosh(2\pi m)$ into  the expressions in 
 \rf{Fnp-int} and \rf{Wnp-int}  
\begin{align}\notag
& F_{\rmNP}(N,y,m) = \cosh(2\pi m)\, F_{\rmNP}(N,y)\,, 
\\[2mm]
& W_{\rmNP}(N,y,m) = \cosh(2\pi m)\, W_{\rmNP}(N,y)\,. \la{119}
\end{align}
 Let us note that the knowledge of the   free energy  in the   massive $\N=2$  $Sp(2N)$ model may be useful 
  for computing (from  its  derivatives over $\l$ and $m$) some  integrated   4-point correlators 
  by analogy with what was done in
   \ci{Chester:2019jas,Chester:2020vyz,Dorigoni:2021guq,Alday:2021vfb,Dorigoni:2022zcr,
  Chester:2022sqb} 
   using the localization results for $\N=2^*$ models 
  (generalizations of  $\N=4$ theories with massive adjoint hypermultiplets)
  for various gauge groups. To  be able to  obtain in this way   interesting examples of correlators 
  one  should actually  generalize our computation to the case   when mass is given to all 
  $\N=2$ hypermultiplets.
  This is a major complication
compared to the case we treat here  because  the simple single trace structure of the matrix model potential is then spoiled.\foot{In particular,  ref. \ci{Alday:2021vfb}  discussed   integrated correlators  in 
  $\N=4$  $Sp(2N)$  theory  given by derivatives of the free  energy in the   $\N=2^*$  $Sp(2N)$ model.
  Let us note also that  linearised discrete  Toda-like relations  that appeared in the finite $\gym$  discussions 
  in \ci{Dorigoni:2021guq,Dorigoni:2022zcr} suggest possible connection
    to \rf{Toda-int}  and a generalization of our discussion beyond the 't Hooft   expansion
    (i.e. including instanton effects). We thank  S. Chester for a discussion  of this connection.}
     
 \subsection*{String theory interpretation}

The  fact that the free  energy  \rf{11} admits a natural expansion in the inverse string tension,   i.e.  in  
$ {1\ov \sql}$, is  already a strong  check  that this  $\N=2$ model  does indeed admit a dual string theory 
description. The   string  theory 
 interpretation of the $\log \l$ term in $F^{\N=4}_N $ in \rf{4} 
   assumes a particular   choice of  an IR cutoff 
 in  the  volume of the AdS$_5$   space which  represents 
  the leading (supergravity) term in the   on-shell value of 
 the  string  effective action   \ci{Russo:2012ay}. 
 The same should   apply to  the $\log $   term in  \rf{11}    (see 
  \ci{Beccaria:2021ism}  and section \ref{S6}   below). 
  
   The structure of the 
    last  term  in \rf{11}, 
    namely $-\tfrac{\pi^{2}}{2}\frac{N}{\l} =- {\pi \ov 8}  {1\ov \gs}$, 
  suggests that it   should come
    from a  disc    or   crosscup  contribution.\foot{A  naive  expectation  is that the string partition function 
  should scale  as  the  volume of AdS$_5$ and  should  then be  always proportional to $\log \l$.
  This is indeed so  in the maximally supersymmetric AdS$_5 \times S^5$ case 
  (where the 1-loop or torus string correction  explains the shift  of $N^2$ term  in 
   the coefficient of the $\log \l$ term, cf. section \ref{S6}). 
   However, this expectation   should  somehow  fail in the  less  supersymmetric cases 
    involving   singular orbifold/orientifold   projections of  AdS$_5 \times S^5$.}
 It is  surprising   that  once   the $\N=2$ free energy is expressed in terms of $\l'$, 
 there are no further  non-trivial contributions 
 of higher order in  expansion in small $\gs\sim {1\ov N} $. This   suggests
 an 
 analogy with   a  non-renormalization 
 of certain protected  quantities (receiving  contributions  only from  few leading orders 
  in perturbation theory)    as it  happens in some models   with extended supersymmetry. 
  
  The   redefinition of $\l^{-1} $  by $1/N$ term in \rf{12}  may be  related to the issue of how one 
  actually compares  gauge theory  to dual string theory, e.g.,     which is the proper 
  definition of   string tension in terms of the  't Hooft coupling $\l$ (and $N$). 
  The
    fact that gauge-theory answer \rf{11}  takes a very simple form when  expressed  
     in 
    terms of $\l'$ rather than $\l$  strongly suggests that 
    it is $\sqrt{\l' }\ov 2\pi $ that  should   be identified  here  with  the string tension.  
    The redefinition $\sqrt \l \to \sqrt {\l'} = { \sqrt{ \lambda} / \sqrt{1 + \betat { \l\ov N}}}$
    may be representing a resummed contribution  of (some of) the open-string  sector  (disc/crosscup)  corrections.  
    
    A test of this would require  a direct computation of the free energy  on 
     the string theory side 
    showing the absence of  all higher   order  $\OO(\gs^n)$         corrections   beyond the ones given in \rf{11}. 
 While this appears beyond our  reach at the moment, in section \ref{S6} 
 we will discuss  the   string theory  derivation 
  of the  subleading in $N$ coefficients 
 of the $\log \l$  terms in the $\N=4$  \rf{4}   and the  $\N=2$  \rf{11}   $Sp(2N)$ models. 
  
  In the  $\N=4$ SYM theory with a gauge group $G$, the coefficient of the  $\log \l$ term in the 
   free energy $F^{\mathcal N=4}$ obtained from the localization matrix model 
  is   proportional to the   conformal anomaly  a-coefficient   given  by  $\four \dim G$  \ci{Beccaria:2021ism}.  
 For example, in the  $SU(N)$  case  the latter  is equal to ${1\ov 4}(N^2- 1)$. While 
   the $N^2$ term comes from  the classical   (supergravity)  part of  string action evaluated on the \adss   vacuum
   (with all $\a'$ corrections vanishing), 
   the additional $(-1)$ term  originates from the one-loop (torus)  string correction  which turns out  to be 
   due  to  the ``massless''   (supergravity)  modes  only  \ci{Beccaria:2014xda}.
   
   As we    demonstrate  in section \ref{S6-1},   a  similar argument explains  the value of the subleading term  in the 
  a-anomaly   or the coefficient  of $\log \l$   in \rf{4} in the $Sp(2N)$   $\N=4$  model   dual to 
   type IIB   string theory on AdS$_5{\times}  \mathbb{RP}^5$ \ci{Witten:1998xy}.
   Here   $\dim G=  N (2N + 1) = 2 (N+\four)^2 - {1\ov 8}$. The shift $N \to N + \four$ 
   may be interpreted as  being due to the change  of the D3-brane  charge  in  the presence of   O3-plane 
    \ci{Blau:1999vz,Giombi:2020kvo}. Then   the remaining  constant   shift   ($-{1\ov 8}$)
     comes again from the 1-loop contribution of the  
     supergravity  modes only.

    In section \ref{S6-2}  we    present an argument that  explains the origin of the order  $O(N)$  term in the coefficient  $N^2 + N + {3\ov 16}$
    of  the $\log \l$  term 
 in  the  free energy \rf{11} of the $\N=2$   $Sp(2N)$ model. 
  However,    it remains a challenge    to reproduce its precise  coefficient  due to currently   insufficient  knowledge   of  
  RR 5-form dependent terms in the D7-brane action in type IIB  background. 
  
  \
  
 The rest of the paper is organized as follows. In section~\ref{S2}, we analyze
  the free energy of the $\mathcal N=2$ $Sp(2N)$ theory. 
  We use localization matrix model to show that it satisfies the Toda lattice equation. 
  We exploit this equation to derive the strong coupling expansion of the free energy and study its properties.
   In section~\ref{S3}, we repeat the same analysis for the circular BPS Wilson loop. 
   The  dual  string theory  interpretation of the  strong-coupling expansions
derived on the gauge theory side is discussed in section~\ref{S6}. 
 
\section{Partition function of  $\N=2$ $Sp(2N)$ theory \la{S2}}
 
In this section, we derive the  large $N$ expansion of the  partition function     of the $\N=2$ superconformal theory
 with   $Sp(2N)$ gauge group  defined on four-sphere. We also find  its  generalization to the case of non-zero mass of fundamental hypermultiplets. 
 
\subsection{Matrix model representation  \la{S2-1}}

For a generic $\N=2$ $Sp(2N)$ theory with hypermultiplets in the fundamental, adjoint, and antisymmetric representations, the localization approach  can be applied to express the partition function  on  a  unit-radius 
 four-sphere   as a matrix integral \ci{Pestun:2007rz}
\ba
\la{4.1}
Z_{\mathfrak{sp}(2N)}= e^{-F_N^{\N=2}}  &= \int  \mc DX\,e^{-  \frac{8\pi^{2}}{g_{{\rm YM}}^{2}}\tr X^{2} }\, |Z_\text{1-loop}(X)|^{2}|Z_\text{inst}(X)|^{2}\,.
\ea
Here the integration goes over $2N \times 2N$ matrices $X$  belonging to the Lie algebra $\mathfrak{sp}(2N)$.  They 
describe zero modes of a scalar field  (from $\N=2$ vector multiplet) on the sphere and have a general form
\begin{align}\label{X}
X = \left[\begin{array}{cc} A & B \\ C & -A^t\end{array}\right]= \sum_{a=1}^{2N^2+N} T^a X^a  \ , 
\end{align}
where $A^t$ denotes a transposed matrix, 
$B^t=B$ and $C^t=C$.  The second relation in \re{X} defines a decomposition of $X$ over the generators of the fundamental representation of  $Sp(2N)$ normalized as $\tr(T^a T^b) = \ha \delta^{ab}$. The integration measure in \re{4.1} is  then $\mc DX=\prod_{a=1}^{2N^2+N} dX^a$. 

In a standard manner, the matrix integral in \re{4.1} can be reduced to an integral over the eigenvalues of  $X$.  By virtue of \re{X}, they take the form $\{\pm x_1,\dots,\pm x_N\}$ leading to
\begin{align}\label{measure}
\mc
 DX=\frac{1}{N!}  
 \prod_{n=1}^{N}   dx_n \, x_n^2   
 \prod_{1\le n<m\le N}(x_{n}^{2}-x_{m}^{2})^{2} 
 \ . 
\end{align}
The functions $Z_\text{1-loop}(X)$ and $Z_\text{inst}(X)$ in \re{4.1} describe the one-loop perturbative correction and the contribution of instantons, respectively. The latter runs in powers of $\exp(-{8\pi^{2}N\ov \l})$ and  is exponentially suppressed at large $N$ and fixed 't Hooft coupling
$\lambda=\gym^2 N$. In what follows we neglect the instanton contribution and put $Z_\text{inst}(X)=1$.
\foot{The instanton contribution is important,  however,  when addressing the large $N$ expansion at fixed $\gym^2$.
Notice also that the absence of instanton corrections at large $N$ implicitly assumes that 
that the integration over instanton moduli does not spoil the instanton action exponential suppression in the $\gym\to 0$ limit. 
This assumption was explicitly checked in $\N = 2$ SQCD  in \cite{Passerini:2011fe}.}

In the $\mathcal N=2$ theory with a number  of hypermultiplets in the fundamental ($n_F)$, adjoint $(n_{\rm adj})$, and antisymmetric ($n_A$) representations,
the one-loop perturbative function $Z_\text{1-loop}(X)$ has the following expression in terms of the eigenvalues of the matrix $X$   
\cite{Fiol:2020bhf}
\ba
\la{4.2}
|Z_\text{1-loop}|^{2} &= \frac{\prod_{n<m}^{N}
[H(x_{nm}^{+})]^2 \, [H(x_{nm}^{-})]^2\prod_{n=1}^{N}[H(2x_{n})]^{2}}
{\prod_{n<m}^{N}[H(x_{nm}^{+})]^{2n_{\rm adj}+2n_{\rm A}}\,[H(x_{nm}^{-})]^{2n_{\rm adj}+2n_{\rm A}}\, 
\prod_{n=1}^{N}[H(2x_{n})]^{2n_{\rm adj}}[H(x_{n})]^{2n_{\rm F}}}
\ea
Here $x^{\pm}_{nm}=x_{n}\pm x_{m}$ and $H(x)$ is expressed in terms of Barnes $G$-functions
\ba\label{H-exp}\notag
H(x) & \equiv   \prod_{k=1}^{\infty}\Big(1+\frac{x^{2}}{k^{2}}\Big)^{k}\, e^{-\frac{x^{2}}{k}}
\\
&
 =  e^{-(1+\gamma)x^{2}}G(1+ix)G(1-ix) = \exp\Big[\sum_{n=1}^{\infty}\frac{(-1)^{n}}{n+1}\zeta_{2n+1}x^{2(n+1)}\Big]\,,
\ea
where $\zeta_{2n+1}\equiv \zeta(2n+1)$ is a Riemann zeta function value.

In this paper  we shall consider 
 the special  $\mathcal N=2$ superconformal  
   theory with the field content 
\be\label{n's}
n_{\rm adj}=0, \qquad\quad n_{\rm F}=4, \qquad\quad n_{\rm A}=1\,,
\ee
for which  the beta-function vanishes.  
Then  \re{4.2} simplifies to\foot{Note that 
 the exponential factors $e^{-\frac{x^{2}}{k}}$ entering the definition \rf{H-exp} of  $H(x)$ (and ensuring  the convergence of the infinite product there)    cancel out  in the ratio of $H-$functions in \rf{4.6}.}
\ba
\la{4.6}
|Z_\text{1-loop} (X)|^{2} \equiv e^{-S_{\rm int}(X)} &= \prod_{n=1}^{N} \frac{ [H(2x_{n})]^2}
{ [H(x_{n})]^{8}} =  \prod_{n=1}^{N}
  \prod_{k=1}^{\infty}{ \big(1+\frac{4x_n^{2}}{k^{2}}\big)^{2k}\ov \big(1+\frac{x_n^{2}}{k^{2}}\big)^{8k}}
    \ . \ea
Substituting this expression into \re{4.1} and setting  $Z_{\rm inst}=1$, we obtain an integral representation for  the partition function of  this 
 $\mathcal N=2$ $Sp(2N)$  model. We will use it 
  to derive the $1/N$ expansion of the free energy $F_N^{\N=2}$.  

  In general,  the partition function of a gauge theory on the sphere suffers  contains ultraviolet divergences
  and requires regularization.  In particular, 
 the form in which $|Z_\text{1-loop}(X)|^{2}$   directly appears from gauge theory calculation
(before the regularization   introduced in  \cite{Pestun:2007rz} leading to  finite $H(x)$ factors)  is 
\ba
\la{666}
|Z^{\rm (bare)}_\text{1-loop}(X)|^{2} \sim  \rr^{1/ 6}  \prod_{n=1}^{N}
  \prod_{k=1}^{\infty}{ \big(k^2 \rr^{-2} +{4x_n^{2}}\big)^{2k}\ov \big(k^2 \rr^{-2} 
  +  {x_n^{2}}\big)^{8k}}
\ , 
\ea
where we restored  the  dependence on the radius $\rr$  of $S^4$
(which  enters also  the Gaussian  action   in \rf{4.1} as 
$\frac{8\pi^{2}\rr^2}{g_{_{\rm YM}}^{2}}\tr X^{2}$).
The  factor $\rr^{1/6} $  stands for the contribution of the ``massless''  multiplets (not depending on $x_n$) 
 which should be   taken  into account  in order  for the r-dependence of  the free  energy 
 $F_N^{\N=2}= -\log Z_{\mathfrak{sp}(2N)}  =  4 \aa \log \rr + ... $    to be   consistent with the 
  value of the \aa-coefficient of the conformal anomaly $\aa= \ha N^2 +\ha  N - {1\ov 24}$
  of   the $\mathcal N=2$ $Sp(2N)$  theory
  (see   Appendix A in \ci{Beccaria:2021ism}).\foot{The factor  of $\rr$ coming from the infinite product  of   ``massive''   modes in \rf{666} is 
  $[\rr^{ 12 \zeta(-1) }]^N = \rr^{-N}$. After the extraction of this factor the dependence 
 of the integrand in \rf{4.1}  on $\rr$  is only through $\rr  x_n$  and thus 
 the extra $\rr$ dependent factor comes  just  from the measure, i.e. is the same as in the $Sp(2N)$ 
 $\N=4$  SYM (Gaussian  matrix model)  case:  $Z_{\N=4} \sim \rr^{-\dim[ {Sp}(2N)]}=  \rr^{-N(2N+1)}$. As a result,  the  dependences of $Z$  on $\l$ and on $\rr$ are  a priori correlated 
 only in the Gaussian model, i.e. in the $\N=4$ SYM case.}
 
The relations \rf{4.1} and \rf{4.6}  can be generalized to a more complicated case of the 
$\mathcal N=2$ $Sp(2N)$ model with the 4  fundamental hypermultiplets having a nonzero mass $m$. Introducing this 
 mass parameter amounts to the replacement the function $|Z_\text{1-loop}|^{2}$  in \re{4.1} by  \cite{Pestun:2007rz}\foot{Here we set again the radius $\rr=1$;
in general, the dependence on $m$ is through the dimensionless combination $m \rr$.}  
\ba\label{Zm}
|Z_\text{1-loop} (X,m)|^{2} &= \prod_{n=1}^{N} \frac{ [H(2x_{n})]^2}
{ [H(x_{n}+m)\ H(x_{n}-m)]^{4}}\,.
\ea
Note that  for  $m \neq 0$ 
 the exponential  convergence factors   in \rf{H-exp}   that ensure the finiteness  of each of  the functions 
 $H$ no longer cancel automatically  in \rf{Zm} giving 
\be \la{mmm}
 \prod_{n=1}^{N} \exp \sum^\infty_{k=1} \Big[ 2 { (2x_n)^2\ov k}  
 - 4 { (x_n+m)^2\ov k}   -   4 { (x_n-m)^2\ov k}   \Big]
 = \exp\Big( - 8 N  m^2  \sum^\infty_{k=1} {1\ov k}  \Big)\ . 
\ee 
The unregularized (bare)
expression  $ |Z^{\rm (bare)}_\text{1-loop} (X,m)|^{2} $ as it originates from the calculation of one-loop 
determinants   in  gauge theory on $S^4$  has the form like in \rf{666}. Namely, it does not  have  these exponential factors  and thus  contains the  logarithmic divergence \rf{mmm} 
proportional to $N m^2$.  

Indeed, 
such logarithmic divergence 
 appears in general  in the partition function of a massive hypermultiplet
defined in  {\it curved} 4-space.\foot{One 
finds (using, e.g.,  proper-time cutoff)  that  $F_\infty= - {1\ov 3 (4\pi)^2 } \int d^4 x\,  \sqrt g\,  R \, m^2 \log \L_{\rm UV}$
where  this divergent contribution comes only from the fermions (assuming  that the scalars  
are conformally coupled, i.e. with   ${1\ov 6} R \phi^2$ term added). 
Note that the $m^4 $ logarithmic   (and all 
 power) divergences cancel out due to  supersymmetry  as in flat space.
In particular,  in the case of $S^4$ of radius $\rr$ (with $R=12 \rr^{-2}, \ {\rm vol}(S^4) = {8\ov 3} \pi^2\rr^4 $),  this gives 
 $F_\infty= -  {2\ov 3 }   (m \rr)^2  \log \L_{\rm UV}$. 
 }
 The finite expression \rf{Zm}  obtained   following 
  \cite{Pestun:2007rz}, i.e.   containing the  factor \rf{mmm},  
 corresponds to a special   choice of the UV subtraction scheme.
 That means, in particular,   that   the  coefficient of  the 
 $N m^2$  term in the resulting free energy is, in general,    scheme-dependent (cf. also a discussion of scheme dependence of $F$ in 
 $\N=2^*$  $SU(N)$ theory in \ci{Chester:2020vyz}).\foot{The same applies to a constant  
 ($N$-dependent) term in the free energy:    for  an $\N=2$ 
 gauge theory defined on $S^4$  the free  energy  contains 
  also  the  UV divergent term  related to the  conformal  anomaly,  
  $F_\infty=  4 \aa  \log \L_{\rm UV}$  (for a generic  curved metric $\aa$ 
  is replaced by a combination of  integrals of the two curvature  contractions 
  with the  a- and c-anomaly coefficients). That means that comparing to string theory one would need to  choose a particular IR  regularization scheme that should correspond to a particular 
  UV regularization  on the  gauge theory side.}
  
To summarize, for  $m\not=0$
the partition function of the $\mathcal N=2$  $Sp(2N)$  model is 
\begin{align}\label{Zm-int}
Z_{\mathfrak{sp}(2N)}{(m)} =  e^{-F_N^{\N=2}(\lambda,m) } =  \int  \mc DX\,e^{-\frac{8N\pi^{2}}{\lambda}\tr X^{2}-S_{\rm int}(X,m)} \,,
\end{align}
with  the integration measure  defined in \re{measure}  and the  interaction action    (expanded in  mass parameter $m$) given by
\ba\label{Sint}
S_{\rm int}(X,m) = -\log |Z_{\rm 1-loop}(X,m)|^{2} 
 = S_{\rm int}^{(0)}+m^{2}S_{\rm int}^{(1)}+m^{4}S_{\rm int}^{(2)}+O(m^6)\,.
\ea 
Taking into account \re{Zm} and \re{H-exp} we find  that 
\ba 
\la{4.8}
& S_{\rm int}^{(0)}  
 = 4\sum_{k=1}^{\infty}\frac{(-1)^{k+1}}{k+1}\zeta_{2k+1}\,(4^{k}-1)\ \tr X^{2(k+1)},
\\
& S_{\rm int}^{(i)}  = 
4\,\sum_{k=1}^{\infty}\frac{(-1)^{k}}{k+1}\zeta_{2k+1}\,\binom{2k+2}{2i}\ \tr X^{2(k+1-i)}\,, \ \ \ \    i\ge 1\ ,  \la{4.7}
\ea
where $\tr X^{2(k+1)}= 2 \sum_{n=1}^N x_n^{2(k+1)}$.
 
\subsection{Large $N$ expansion of  free energy \la{S2-2}} 

We observe that  in the planar limit,
for $N\to\infty$ with $\lambda =\text{fixed}$, the potential in the matrix integral \rf{Zm-int} is dominated by a Gaussian term. Because  
the contribution of the interaction term $S_{\rm int}(X)$ to the partition function \re{Zm-int} is suppressed by a factor of $1/N$,
 the free energy of the $\mathcal N=2$ model 
$F_N^{\mathcal N=2}(\lambda,m) = - \log Z_{\mathfrak{sp}(2N)}(m)$ coincides in the planar limit with the free energy of  the $\mathcal N=4$ SYM 
theory
\begin{align}\la{1v}
e^{-F_N^{\mathcal N=4}(\lambda)} =  \int  \mc DX\,e^{-\frac{8N\pi^{2}}{\lambda}\tr X^{2}}\,.
\end{align}
This suggests to define the free energy difference 
\begin{align}\la{2v}  
\Delta F(\lambda;N,m) &= F_N^{\mathcal N=2}(\lambda,m) - F_N^{\mathcal N=4}(\lambda) \ .  
\end{align}
Its large $N$  expansion   starts with  order $N$ term and runs in powers of $1/N$
\begin{align}\label{Delta-F}
\Delta F(\lambda;N,m) =  N \FF_1(\lambda,m) +  \FF_2(\lambda,m) +{1\over N} \FF_3(\lambda,m) + \OO({1\ov N^2})\ . 
\end{align}
Here the corresponding coefficient functions $F_n(\lambda,m)$  
are given by cumulants of $S_{\rm int}(X;m)$ in the  Gaussian matrix model. 
For example, from \rf{2v} the leading term  in \re{Delta-F} is given by  
\begin{align}\label{F1-exp}
\FF_1(\lambda, m) = \lim_{N\to\infty}{1\over N} \vev{S_{\rm int}(X;m)}
= \FF_1(\lambda)+m^2 \, \FF_1^{(1)} (\lambda)+m^4 \, \FF_1^{(2)} (\lambda)+\dots \,
 \ , 
\end{align}
where $\FF_1(\lambda)\equiv \FF_1(\lambda, 0)$ corresponds to the massless theory and the angular brackets denote an
 average in the  Gaussian matrix model. 

Replacing $S_{\rm int}(X;m)$ in \rf{F1-exp}
with its small $m$ expansion \re{4.7}, we get an expression for $\FF_1^{(i)}(\lambda)$ as an infinite sum of terms proportional to $\vev{\tr X^{2(k+1-i)}}$. 
\footnote{Higher order terms in the $1/N$ expansion, \cf (\ref{Delta-F}), involve connected correlators (cumulants) of product of traces $\vev{\tr X^{k_{1}}\cdots \tr X^{k_{n}}}$.}
These expectation values can be computed in the  Gaussian $Sp(2N)$ model  in the large $N$ limit using the technique developed in \cite{Beccaria:2021ism}. Replacing the Riemann function  value $\zeta_{2n+1}$ in \re{4.7} 
with its integral representation, one can resum the series in \re{4.7} to obtain the leading term of the small $m$ expansion \re{F1-exp} as \ci{Beccaria:2021ism}
\begin{align}\label{F0}
\FF_1(\l) &= \frac{4}{\sql}\int_{0}^{\infty}dt\frac{e^{2\pi t}}{(e^{2\pi t}-1)^{2}}\frac{3t\sql-8J_{1}(t\sql)+J_{1}(2t\sql)}{t^{2}}\,,
\end{align}
where $J_1$ is a  Bessel function. 

For $m\neq 0$, the expansion  in \re{F1-exp} can be summed up  in a similar manner  to all orders in $m^2$ to give 
\be
\la{8.20}
\FF_1(\l, m) = \FF_1(\l) + \frac{64}{\sql}\,\int_{0}^{\infty}dt\, \frac{e^{2\pi t}}{(e^{2\pi t}-1)^{2}}\Big[\frac{J_{1}(t\sql)}{t^{2}}\sin^{2}(m\pi t)-\frac{1}{2}\sql\, m^{2}\pi^{2}t\Big].
\ee
The relations \re{F0} and \re{8.20} define the leading large $N$ correction to the difference free energy \re{Delta-F} in the 
massless and massive  $\mathcal N=2$ theory, respectively. 
They  are valid for  an arbitrary 't Hooft coupling $\lambda$.  
At weak coupling, it is straightforward to expand $\FF_1(\l)$ and $\FF_1(\l, m)$ in powers of $\lambda$.   

Here  we will concentrate  on studying the difference free energy \rf{2v} at strong coupling.
At strong coupling, the expansion of the functions  $\FF_k(\l)$ in \rf{Delta-F}  runs in powers of  the two parameters $1\ov \sqrt{\lambda}$ and $e^{-\sqrt{\lambda}}$. 
In particular, $\FF_1(\l, m) $ can be split into the sum of the two terms   
\begin{align}\la{227}
\FF_1(\l,m) &= \FF_{1,\rmP}(\l,m) + \FF_{1,\rmNP}(\l,m) \,.
\end{align}
The first (``perturbative'') term is given by a  series in $1\ov \sqrt{\lambda}$ and the second 
(``nonperturbative'') one is  a sum of  terms containing powers of exponentially small 
factors  $e^{- \sqrt{\lambda}}$. 

For  $m=0$ one finds from \rf{F1-exp} and \rf{F0} (see \cite{Beccaria:2021ism})
\ba\label{F1-P+NP}
&\FF_1(\l) \equiv \FF_1(\l,0)= \FF_{1,\rmP}(\l) + \FF_{1,\rmNP}(\l) \,,
\ea
where 
\ba \la{829}
& \FF_{1, \rmP}(\l) =\te  \frac{\log 2}{2\pi^{2}}\,\lambda-\frac{1}{2}\log \l+\big(\log\pi+\frac{7}{3}\log 2+\frac{3}{2}-12\log\mathsf{A}\big) -\frac{\pi^{2}}{2\l}\ ,
\\[2mm]
& \FF_{1, \rmNP}(\l) =\te   \frac{8\sqrt 2}{\pi^{3/2} }\l^{-1/4}e^{-\sql}\Big(1+\frac{23}{8\sql}+\frac{153}{128\l}-\frac{435}{1024\l^{3/2}}
+\cdots\Big)+\mc O(e^{-3\sql})\ , \la{8.29}
\ea
with $\mathsf{A}$ being the  Glaisher's constant. 
Remarkably,  the strong coupling expansion of $F_{1, \rmP}(\l)$ contains only a finite number of terms and terminates at order $O({1/\lambda})$.

In the mass-deformed $\mathcal N=2$ theory    we  found  that the two  terms in \rf{227}  are  
 \begin{align}\la{230}
& \FF_{1,\rmP}(\l,m) = \FF_{1, \rmP}(\l)-\Big[4\log\l+4\big(1+2\gamma_{\rm E}-2\log(4\pi)\big)+\tfrac{16\pi^{2}}{3\l}\Big] m^2-\tfrac{16\pi^{2}}{3\l}m^4\,,
\\[2mm]
& \FF_{1,\rmNP}(\l,m) = \tfrac{8\sqrt 2}{\pi^{3/2}}\,\l^{-1/4}\,e^{-\sql}\,\cosh(2\pi m) +\dots\,,\la{231}
\end{align}
where $\gamma_{\rm E}$ is the 
Euler's constant\foot{As was discussed above (cf. \rf{mmm}), 
 the constant coefficient of the $m^2$  term in 
\rf{230} is renormalization scheme dependent.} 
and dots in \rf{231}  denote terms suppressed by $1/ \sqrt\lambda$ or extra  $e^{-\sql}$ factors as in \rf{8.29}. 
Note that the small $m$ expansion of $\FF_{1,\rmP}(\l,m)$ terminates  at order $m^4$ whereas the dependence of $\FF_{1,\rmNP}(\l,m)$ on 
  $m$ enters through the   function  $ \cosh(2\pi m)$.\foot{Equivalently,   we get 
  $\FF_{1,\rmNP}(\l,m) \sim \ha(e^{-\sql + 2\pi m} +e^{-\sql -2\pi m})+\dots$   
  hinting at possible world-sheet instanton interpretation.}

The same techniques can be applied to compute  subleading terms 
 in the large $N$ expansion of \re{Delta-F}, 
    but  calculations become cumbersome as one goes to higher orders in $1/N$. 
    It turns out,  however, 
     that the resulting expressions for the functions $\FF_{n\ge 2} (\lambda,m)$  can  all  be  effectively expressed  
      in terms of   the leading large  $N$ function $\FF_1(\lambda, m)$.
      In particular,  for $m=0$, one finds \ci{Beccaria:2021ism}  
\ba\notag \la{88}
& 
 (\FF_2(\lambda)){}'  = \tfrac{1}{4}(\lambda \FF_1(\lambda))''  -\tfrac{1}{4}\l\big[(\lambda\, \FF_1(\lambda))''\big]^{2} \ , 
 \\[2mm]
& \FF_{3}(\lambda)  =  
\tfrac{1}{48}\l^2\big( \l \FF_1(\lambda)\big)'''  - \tfrac{1}{16}\l^2\big[ \big(\l \FF_1(\lambda)\big)''\big]^{2}  +\tfrac{1}{24}\l^{3}\big[ \big(\l \FF_1(\lambda)\big)''\big]^{3}\ ,  \ \ \   etc.\ ,
 \ea
where prime denotes a derivative with respect to the 't Hooft coupling $\lambda$. 
The origin of these relations  will be  explained in the next subsection. 

\subsection{Free energy from   Toda  lattice equation \la{S2-3}}  

As discussed  above, the 
 partition function of the $\mathcal N=2$ gauge  theory under consideration can be represented as 
the  partition function of the $Sp(2N)$ matrix model \re{Zm-int}. 
It is well-known that for  a generic matrix model with the potential given by a 
sum of single trace terms  with
 arbitrary coefficients, $V(X)=\sum_k t_k\,  \tr X^k$,
  its partition function satisfies nontrivial relations that define an integrable Toda-like hierarchy. Such relations have been  
studied in past in the context of a unitary matrix model \cite{Gerasimov:1990is,Martinec:1990qg,Alvarez-Gaume:1991xsn,Morozov:2009uy}.  
Similar relations  apply also  to the $\mathfrak{sp}(2N)$ matrix model
with a  generic single-trace potential. 

The matrix integral \re{Zm-int} corresponds to a particular choice of  the coefficients 
 in a generic potential $V(X)$ of the $\mathfrak{sp}(2N)$ matrix model. Namely, the coefficient of 
the Gaussian term $\tr X^2$  is proportional to 
the effective  coupling constant
\begin{align}\label{y}
y={(4\pi)^2\over \gym^2} = {(4\pi)^2N \over \lambda} \,, 
\end{align}
whereas the coefficients in front of the  other single trace terms  $\tr X^{2k}$ (with $k\ge 2$) are 
uniquely fixed  by   the localization representation, see Eqs.~\rf{4.8} and \rf{4.7}. 

To simplify the notation,  let  us 
denote the partition function \re{Zm-int} 
as $Z_N(y)\equiv Z_{\mathfrak{sp}(2N)}$. In general, it is a function of $N$ and the 
 inverse coupling constant \re{y}, with the  dependence on $m^2$  tacitly assumed.
Repeating the analysis of \cite{Gerasimov:1990is,Martinec:1990qg,Alvarez-Gaume:1991xsn}, we find that it satisfies  the following Toda lattice equation 
\begin{align}\label{Toda}
\partial^2_y \log Z_N(y) = {Z_{N+1}(y)\, Z_{N-1}(y)\over Z^2_{N}(y)}\,.
\end{align}
The expression on the right-hand side involves the partition functions defined for the same value of $y$ and with  $N$ shifted by $\pm 1$.\foot{Note that using  $\lambda={(4\pi)^2N\ov y}$ instead of $y$ as an argument of $Z_N$   would   make
this equation rather  cumbersome.}

The equation  \re{Toda} is supplemented by 
 the boundary conditions 
 \be Z_{N=-1}(y)=0  \ , \qquad  \ \ \ Z_{N=0}(y)=1 \ . \la{bc}\ee
 $Z_{N=1}(y)$  gives   the partition function of the $\N=2$ $Sp(2)$ theory
\begin{align}
Z_{N=1}(y) =  \int_{-\infty}^\infty dx\, x^2 \,e^{-y \,x^2-S_{\rm int}(X,m)} \,,
\end{align}
where $2\times 2$ matrix $X$ has the eigenvalues $\{x,-x\}$. 
Here $S_{\rm int}(X,m)$ is given by Eqs.~\rf{Sint} -- \rf{4.7} with $\tr X^{2k} = 2x^{2k}$.
 
 Applying the Toda equation \re{Toda} 
  recursively, we can express the partition function $Z_N(y)$
   in terms of only  {\it one}  function $Z_1(y)$ and its derivatives.
   The  general solution is \cite{forrester2001application}  
\be
\la{5.42}
Z_N(y) = \det \| \partial_y ^{j+k} Z_1(y) \|_{_{0\le j,k\le N-1}}
=\det \left[\begin{array}{llll} Z_1 & Z_1' & \dots & Z_1^{(N-1)} \\ Z_1' & Z_1'' & \dots & Z_1^{(N)} \\ \vdots & \vdots  &  
& \vdots \\ Z_1^{(N-1)} & Z_1^{(N)} & \dots & Z_1^{(2N-2)} \end{array}\right]\ , 
\ee
where $Z_1^{(k)}\equiv \partial_y^k Z_1(y)$. 

It  should be  noted that the 
 Toda equation \re{Toda} does not depend on particular  values of the  coefficients 
 in  the interaction term $S_{\rm int}(X,m)$.  
 For example, 
  it   should hold   both in  $\mathcal N=2$ and $\mathcal N=4$  $Sp(2N)$   models.  
In  the latter case  $S_{\rm int}(X,m)=0$  and   
 the partition function can be computed  directly  from the matrix integral 
  for an arbitrary $N$ using  the  orthogonal polynomial technique \ci{mehta}
\ba\label{Z4}\notag
& Z_N^{\mathcal N=4}(y) =e^{-C^{\mathcal N=4}_N}\, y^{-\frac{1}{2}N(2N+1)}\,,
\\
& C^{\mathcal N=4}_N=  \log\frac{G(N+1)\, G(N+\frac{3}{2})} {G(\frac{3}{2})}\ , 
\ea
where   
$G$ is Barnes function. One can verify that this  $Z_N^{\mathcal N=4}(y)$  indeed 
 satisfies \re{Toda} and \re{5.42}. 
The  free energy in the $\mathcal N=4$ $Sp(2N)$ theory 
  is thus 
\ba\notag
F_N^{\mathcal N=4}(\lambda) & \equiv   -\log Z_N^{\mathcal N=4} (y) 
\\[2mm]
& = \
    \tfrac{1}{2}N(2N+1)\log y  +C_N^{\mathcal N=4} \no 
\\[2mm]
    &=  -\tfrac{1}{2}N(2N+1)\log \l  + \tfrac{1}{2}N(2N+1) \log [(4\pi)^2 N]
     +C_N^{\mathcal N=4}\ , \label{F-N=4}
\ea
where in the last relation we replaced $y$ with $\l$    according to  its definition in  \re{y}.

In the $\mathcal N=2$ model we have from \rf{2v} 
\begin{align}\la{240} 
Z_N^{\mathcal N=2}(y) = \exp\left( -F_N^{\mathcal N=4}(\lambda)-\Delta F(\lambda;N,m) \right)\Big|_{\l={(4 \pi)^2 N\ov y}}\,,
\end{align}
where $F_N^{\mathcal N=4}(\lambda)$ is given by \re{F-N=4}. Substitution of this expression into \re{Toda} leads to a nontrivial equation  for 
the function $\Delta F(\lambda;N,m)= \Delta F({(4 \pi)^2 N\ov y};N,m) $. 
 Using the general expression \re{Delta-F}  for $\Delta F(\lambda;N,m)$ and
expanding both sides of \re{Toda} at large $N$ with fixed $\lambda$  we get
to the  leading order in $1/N$  
\begin{align}\label{6.12}\notag
\FF_2'(\lambda,m) {}& = \tfrac14 \GG(\lambda,m)\big[1-\lambda \GG(\lambda,m)\big]\,,   
\\[2mm]
\GG(\l,m) {}& \equiv  \partial_\lambda^2\big(\lambda \FF_1(\lambda,m)\big)\,,  
\end{align}
where $\FF_1(\lambda,m)$   is given by \rf{8.20}.  
At the   next order in $1/N$,  from  \re{6.12} we  find the following  differential equation for $\FF_3$
\begin{align}\label{pdeF3} 
 \FF_3(\l,m )-2 \l  \FF_3'(\l ,m) {}& =\te \frac{3}{16} \l ^2 
 \GG ^2-\frac{5}{24} \l ^3 \GG ^3-\frac{1}{16} 
\l ^2 \GG {'}
 + \frac{1}{4} \l ^3  \GG ^2
\GG {'}-\frac{1}{24} \l ^3 \GG {''}  \,,
\end{align}
where $\GG'=\partial_\lambda  \GG(\lambda,m)$, etc.
Its general solution is 
\ba
\la{6.14}
\FF_{3}(\l,m) &= 
c \,\sqrt{\l }+\tfrac{1}{48} \l ^2 \Big(-3  \GG^2(\l,m )+2 
\l   \GG^3(\l,m )+\GG {'}(\l,m )\Big)\, ,
\ea
where $c $ is an integration constant. The value of $c$ can be found using the weak-coupling expansion of $\FF_3(\lambda)$. As this expansion runs in powers of  $\lambda$, 
it can not contain  an  $\OO(\sqrt\lambda)$ term. Using 
 \re{6.14} this then leads to the conclusion that  $c=0$. 

To summarize, we find that 
\begin{align}\notag\label{FF's}
\FF_2(\lambda,m) {}& = \tfrac14 \int_0^\lambda d\lambda\, \GG(\lambda,m)\big[1-\lambda \, \GG(\lambda,m)\big]\,,
\\[1.2mm]
\FF_{3}(\l,m) {}&= \tfrac{1}{48} \l ^2 \Big[-3  \GG^2(\l,m )+2 
\l  \,  \GG^3(\l,m )+\GG '(\l,m )\Big]\,,
\end{align}
where $\GG(\l,m)$ is given by \re{6.12}. 

Expanding both sides of \re{Toda} to higher orders in $1/N$ we can express all the subleading coefficient functions in \re{Delta-F} in terms of $\GG(\l,m)$.
In distinction to $\FF_2(\lambda,m)$, the functions $\FF_k(\l,m)$ with $k\ge 3$ depend locally on $\GG(\l,m )$ and a finite number of its derivatives.
These relations  
 hold for  an arbitrary mass parameter $m$ (for   $m=0$ they coincide with the relations in \rf{88}). 
 This is again   because of the universality of  the Toda equation \rf{Toda}  that applies to any  single-trace potential. Thus the information about the mass parameter enters only  through {\it one}  function  $\FF_1(\l,m)$  given by \rf{8.20}.

The  expressions \re{FF's}, etc.,  for $\FF_n(\lambda,m)$ (with $n\ge 2$) are valid for an arbitrary  value of 't Hooft coupling. 
Taking into account \re{230} and \re{231}, we can then systematically work out the strong coupling expansion of the free energy \re{Delta-F} to any order in $1/N$.
From 
  \re{6.12}, \re{230} and \re{231} we find that the  strong coupling expansion of $\GG(\l,m )$ is given by 
\begin{align}\notag\label{calG}
\GG(\l,m ) {}& =\GG_{\rmP}(\l,m ) +\GG_{\rmNP}(\l,m ) 
\\
{}& = \Big({\frac{\log 2}{\pi ^2} -\frac{\frac{1}{2}+4 m^2}{\lambda }}\Big)+  \frac{2 \sqrt{2} }{\pi ^{3/2}} \lambda^{-1/4} e^{-\sqrt{\lambda }} \cosh (2 \pi  m)\,  \big(1+ \OO(\lambda^{-1/2})\big)   \,.
\end{align}
Substituting this expression into \re{FF's} we get
\begin{align}\notag
\FF_2(\lambda,m)={}& -\tfrac12\betat^2 \lambda ^2+    \left(4 m^2+1\right)\betat\lambda -(\tfrac34+2m^2)(\tfrac14+2m^2) \log  \lambda +f(m^2)
\\
 {}&+ \tfrac{4\sqrt 2}{ \pi^{3/2}}\,\betat \,\l^{5/4}e^{-\sqrt{\lambda }} \cosh (2 \pi  m)\, \big(1+ \OO(\lambda^{-1/2})\big) \ , \la{r1}
\\[2mm]\notag
\FF_3(\lambda,m) = {}&  \tfrac13 \betat^3 \lambda ^3 -\tfrac12  \left(4 m^2+1\right)\betat^2 \lambda ^2  +   (\tfrac34+2m^2)(\tfrac14+2m^2)\betat \lambda
+\OO(\lambda^0)\ , \la{r2}
\\
 {}&+ \tfrac{4\sqrt 2}{\pi^{3/2}}\,\betat^2 \,\l^{11/4}e^{-\sqrt{\lambda }} \cosh (2 \pi  m)\,  (\tfrac14+ \OO(\lambda^{-1/2})) \ , 
\end{align}
where $b= {1\ov 2 \pi^2} \log 2 $ as in  \re{12}. 

 The expression for $\FF_2(\lambda,m)$ involves the $\lambda$-independent function $f(m^2)$.  It arises as an integration constant of the differential equation \re{6.12} and should be determined independently.\footnote{We are grateful to the authors of \cite{Behan:2023fqq}  for pointing our the missing $O(\lambda^0)$ term in \re{r1} in the previous version of the paper.} 
Namely, the large $\lambda$ expansion of $\FF_2(\lambda,m)$ can be 
derived by combining together the relations \re{FF's}, \re{6.12} and \re{8.20}. Matching it to \re{r1} we found after some algebra (here $\log \mathsf{A}= {1\ov 12} - \zeta'(-1)$)
\begin{align}\label{f-cor}
 \notag
f(m^2) = {}&\frac{1}{3}+\frac{3 \log \pi}{8} +\frac{221 \log 2}{360} -4 \log \mathsf{A} -5 \zeta '(-3)
 \\\notag
{}& +m^2 \Big[ -4 \gamma -\frac{13}{3}+4 \log \pi
   +\frac{20 \log 2}{3}\Big]
\\\notag
{}&+m^4 \Big[ 8 \zeta
   (3)-8 \gamma -\frac{44}{3}+8 \log(4\pi) \Big]   
\\
{}&+ \sum_{p\ge 3} {(-4m^2)^{p}\over p}  \Big[\zeta(2p-1)-\zeta(2p-3) \Big]\,.
\end{align}
The first few terms of the expansion are in agreement with the results obtained in \cite{Behan:2023fqq}.  

 The first and the second lines in \rf{r1} and \rf{r2} define the perturbative and nonperturbative corrections, respectively. The analogous expression for the leading term $\FF_1(\l,m)$ is given by Eqs.~\re{227} -- \re{230}. 

Using  the obtained expressions for $\FF_k(\lambda,m)$ (with $k=1,2,3$)  in  \re{Delta-F} we observe that the strong coupling expansion of the difference free energy  has 
 an  interesting structure
\begin{align}\notag\label{hint}
\Delta F(\lambda;N,m)=  {}&  \te N  \betat\lambda  -\frac12 (\betat\lambda)^2 + \frac1{3N} (\betat\lambda)^3 + \dots
\\
{}& \te + N \tfrac{8\sqrt 2}{\pi^{3/2} }\l^{-1/4}e^{-\sql}\cosh (2 \pi  m)\Big[1+ {\betat \lambda^{3/2}\over 2N} +\frac12\big({\betat \lambda^{3/2}\over 2N}\big)^2+\dots   \Big]\,,
\end{align}
where dots denote subleading corrections. This suggests that the strong coupling expansion can be resummed to all orders in $1/N$. Moreover, 
introducing the new expansion parameter
\begin{align}
\lambda' ={\lambda\over 1+ \betat \lambda/N}\ , \la{lapp}
\end{align}
we find that the relation \re{hint} leads to a remarkably simple expression for $\Delta F(\lambda; N,m)$
\begin{align}\label{hint1}
\Delta F(\lambda;N,m) = N^2 \log (\lambda/\lambda') +   N \tfrac{8\sqrt 2}{\pi^{3/2} }\, {\l'}^{-1/4}e^{-\sqrt{\l'}}\cosh (2 \pi  m) 
+ \dots\ . 
 \end{align}
In the next subsection we  explain how this relation naturally follows from the Toda equation \re{Toda}
and also  comment on the origin of the redefinition  \rf{lapp}.
 
\subsection{Resummation  \la{S2-4}}  

Let us examine the Toda equation \re{Toda} at strong coupling. As was explained above, 
solving  this equation it is advantageous to consider the free energy
 $F=-\log Z_N $ as a function of the inverse coupling $y$  in  \re{y}  rather than of the 
  't Hooft coupling $\l$. 
  
We have seen that the free energy of the $\mathcal N=4$ theory \re{F-N=4} is given by the sum of a  term proportional to $\log y$ (or $\log \l$) 
and a constant. In the $\mathcal N=2$ theory the situation is more complicated --  the difference free energy \re{Delta-F} receives corrections  which are series  in 
    $1\ov \sqrt{\lambda}$ and $e^{-\sqrt\lambda}$. To leading order in $1/N$ they are given by  \re{227}.   Similarly,     the free energy   
 can be in general 
 split into the sum of perturbative and nonperturbative pieces
\begin{align}\la{npp}
F_N^{\mathcal N=2} (y,m)= F_{\rmP}(N,y,m) + F_{\rmNP}(N,y,m) \ , 
\end{align}
where  the second term 
is suppressed by a factor of $e^{-\sqrt\lambda}= e^{-4 \pi \sqrt{N \ov y}}$.

Substituting $Z_N(y)=\exp(-F_N^{\mathcal N=2} (y))$ into \re{Toda} and neglecting 
all $\OO(e^{-\sqrt\lambda})$ corrections on both sides of  the equation
we get the same relation  just for the 
perturbative part of the free energy\foot{For  simplicity,  we will 
  not  explicitly display  the dependence on $m$ in the equations below.}
\begin{align}\label{Toda-P}
\partial_y^2 F_{\rmP}(N,y) = -e^{-\Delta_{_N} F_{\rmP}(N,y) }\,,
\end{align}
where we  introduced the    notation for the second-order finite-difference operator
\begin{align}
\Delta_{_N} F_{\rmP}(N,y) \equiv F_{\rmP}(N+1,y)-2F_{\rmP}(N,y)+F_{\rmP}(N-1,y) \,.
\end{align}
In a similar manner, matching 
the  nonperturbative $O(e^{-\sqrt\lambda})$ terms on both sides of \re{Toda} we obtain 
\begin{align}\la{246}
\partial_y^2 F_{\rmNP}(N,y) = - \Delta_{_N}  F_{\rmNP}(N,y) 
  \ \partial_y^2 F_{\rmP}(N,y)  + \OO(e^{-2\sqrt\lambda})\,.
\end{align}
It is straightforward to extend the analysis to take into account subleading  nonperturbative corrections $\OO(e^{-n\sqrt\lambda})$ with $n\ge 2$. In what follows we restrict consideration to the leading $\OO(e^{-\sqrt\lambda})$ nonperturbative  terms.

The solutions to \rf{Toda-P} and \rf{246} should respect the symmetry of the  Toda equation  \re{Toda}.
As the equation \re{Toda}  is  invariant under an $N$-independent  constant  
 shift  of $y$
 \be   \la{2466}
 y\to y +  c_0 \ , 
 \ee
 the functions $F_{\rmP}(N,y+c_0)$ and $F_{\rmNP}(N,y+c_0)$
satisfy  \rf{Toda-P} and \rf{246} for an arbitrary $c_0$. 
The solutions to  \rf{Toda-P} and \rf{246} are defined up to a contribution of the  zero modes of the operators $\del^2_y$ and $\Delta_{_N}$.  The zero mode solution
$\del^2_y F_{\rm zero}=\Delta_{_N} F_{\rm zero}=0$
  has  the form ($c_i$ are arbitrary constants) 
\be \la{zzz}
F_{\rm zero}(N,y) = c_1+ c_2  N + (c_3+ c_4 N)\, y \ . 
\ee
Taking this into account, we look for the solution to \re{Toda-P} as
\begin{align}\label{3terms}
F_{\rmP}(N,y) = f_0(N) \log\big( y + \kappa \big)  + f_1(N)\,  y\ + f_2(N)\,,
\end{align}
where $N-$independent constant $\kappa$ and  the  functions $f_i(N)$  
are  to be determined.  

The motivation for choosing the ansatz \re{3terms} is twofold. First, it is consistent with the shift symmetry 
\re{2466} and it takes into account the contribution of  the zero mode \rf{zzz}.
Second, for $\kappa=0$ its functional dependence on $y$ matches that of the free energy in the  $\mathcal N=4$ theory and  the leading non-planar correction to the difference free energy in the $\mathcal N=2$ theory, Eqs.~\re{F-N=4}  and  \re{230}, respectively.

While we could  eliminate $\kappa$ in \re{3terms} using the  shift symmetry \rf{2466},  
keeping it non-zero is important in order to correctly reproduce  the $\lambda$-dependent terms in the  $1/N$ expansion of the free energy. 
To see this, we express $y$ in terms of $\l$   according to its definition  in \rf{y}    and expand the first term in \re{3terms} 
at  large $N$ and fixed $\l$  
\begin{align}\label{large-N}
F_{\rmP}(N,y) =- f_0(N) \log \lambda +f_0(N) \log\Big(1+{\lambda\,\kappa\over (4\pi)^2N } \Big) +
 f_1(N)\,  {(4\pi)^2N \over \lambda} +\dots\,,
\end{align}
where dots denote $\lambda$-independent terms. 
Due to planar equivalence between  the $\mathcal N=4$ and $\mathcal N=2$ theories, the expression  \re{large-N} should coincide 
in the leading large $N$, fixed $\l$  limit  with \re{F-N=4}.
 This leads to the  constraints  
 \be\la{77}
  f_0(N) = N^2 +\OO(N)\ , \ \ \ \qquad  f_1(N) = \OO(N^0) \ . \ee
Then, 
it follows from \re{large-N} that all the terms in the large $N$ expansion of $F_{\rmP}(N,y)$  which have 
 the form $\lambda^k/N^{k-2}$ (with $k\ge 1$) arise from the expansion of the  logarithm in the second term. 
 In particular, for $k=1$ we get 
 $\lambda N\kappa/(4\pi)^2$. It should be compared with the analogous term 
$ \lambda N \log 2/(2\pi^2)$
in the expression for the leading non-planar correction $N \FF_{1,\rmP}$, Eqs.~\re{230} and \re{829}. This fixes  the value of   $\kappa$ (cf. \rf{12})
\be \la{977}
\kappa = 8 \log 2 = (4 \pi)^2 \betat \, ,
\ee
where $b$ is the same  as in \rf{12}.

To determine the functions $f_i(N)$ in \rf{3terms} we 
substitute \re{3terms} into \re{Toda-P} and compare the $y$-dependence on both sides to obtain
\ba  
& \Delta_{_N} f_0(N) =2\,,\qquad 
\Delta_{_N} f_1(N) = 0\,,\qquad  \la{248}
\Delta_{_N} f_2(N)= -\log f_0(N)\ . 
\ea
We require that the solution of these equations  should admit a regular 
large $N$ expansion.
Then, the general solution to the first two equations in \re{248}  consistent with  the  constraints   in \re{77} is
\begin{align}\notag
& f_0(N) = (N+N_+)(N+N_-)\,,
\\[2mm] 
& f_1(N) = c_3 \,, \la{250}
\end{align}
where $N_\pm$ and $c_3$ so far are arbitrary  constants. 
The solution of the last equation in \re{248} can be expressed in terms of the Barnes function
\begin{align}\label{c3,c4}
f_2(N) = -\log \big[{G(N+1+N_+)G(N+1+N_-)}\big]+c_1+c_2 N\,,
\end{align}
where the last two terms represent the  zero modes. 

As before, we can fix the values of the parameters in \re{250} and \re{c3,c4} by comparing \re{large-N} with the first few terms of the large $N$ expansion of the difference free energy \re{Delta-F}. To find $c_3$ in \re{250}, it is sufficient to compare $\OO({N\ov\lambda})$ terms in \re{large-N} and  $N \FF_{1,\rmP}$, Eqs.~\re{230} and \re{829}. This leads to
\begin{align}\label{c3}
c_3 =\te -\frac{1}{32}- \frac{1}{3} m^2-\frac{1}{3}m^4\,.
\end{align}
To find  $N_\pm$, we examine the coefficient in front of $\log\lambda$ in \re{large-N} and match it  with  $f_0(N)$ in \re{250}.
To this end, we need the expressions for the first two terms in  \re{Delta-F}. 

We recall that the corresponding functions $\FF_1(\lambda)$ and $\FF_2(\lambda)$ are related to each other through \re{88}. 
According to \re{230} and \re{829}, the $\log \lambda$ term in $\FF_1$ has the coefficient 
\be  q \equiv -\tfrac12-4m^2\ . \ee
Eq.  \re{88} implies that the same term in $\FF_2$ has the coefficient $\tfrac14 q(1-q)$. Combining this with the contribution from \re{F-N=4}, we find the coefficient of  $\log \lambda$ term in $F_{\rmP}$ as 
\begin{align}
\te
f_0(N)=\frac12 N(2N+1)-q N-\frac14 q (1-q)=(N-\frac12 q)\big(N+\frac12(1-q)\big)\,.
\end{align}
Comparing this relation with \re{250} we deduce that  
\begin{align}\la{npm}
N_-=\tfrac14+2m^2\,,\qquad\qquad N_+=\tfrac34+2m^2\,.
\end{align}
Finally, the  constants $c_1$ and $c_2$ in \re{c3,c4} can be determined by matching the  $\lambda$-independent terms in the 
expression for $\FF_1$ in \re{230} and $\FF_2$ in  \re{88}   with \re{3terms}. This leads to
\begin{align}\notag\label{c1c2}
& c_1 =-\tfrac18\log (16\pi)+\log G(\tfrac32) -2 m^2 \log (8 \pi )-8 m^4 \log (4 \pi ) +f(m^2)\,, 
\\[1.2mm]  
& c_2 =-2 \log \pi + 8\log G(\tfrac32) -m^2 (8+8 \gamma_E )\,,
\end{align}
where  $f(m^2)$ is defined in \re{f-cor} and 
the  Barnes function
 $G(\tfrac32)$ can be expressed in terms of  the Glaisher's constant.
 
Combining together the above relations, we arrive at the following remarkably simple 
expression for the perturbative part of the free energy \re{3terms}
\begin{align}\notag\label{resum}
F_{\rmP}(N,y,m)= &(N+\tfrac34+2m^2)(N+\tfrac14+2m^2) \log\big( y +  8 \log 2 \big)  
\\[2mm]
& -\log \left[{G(N+\tfrac54+2m^2)G(N+\tfrac74+2m^2)}\right]
 +c_1+c_2 N+ c_3\,  y \,,
\end{align}
where we explicitly  indicated  the dependence  of $F_{\rmP}$   on $m$.  
The last three terms  in \rf{resum} are the zero modes \rf{zzz}
 of the operators $\del^2_y$ and $\Delta_{_N}$. 
 The corresponding coefficients are given by \re{c3} and \re{c1c2}.

\
 
 Let us now make  few comments.
The expression \re{resum} sums up (perturbative) strong coupling corrections to the free energy  to all orders in $1/N$ and  $1/ \sqrt\lambda$.
Following \re{large-N}, we can expand \re{resum} in $1/N$ and determine the perturbative part of the functions $\FF_k(\lambda,m)$ in \re{Delta-F}. It is straightforward to verify that these functions satisfy \re{88}.

Notice that $N$ and $m^2$ enter the first two terms in \re{resum} in a linear combination  $N+2m^2$.  This means that, up to the contribution \rf{zzz} of the zero modes, the dependence of the free energy on $m$ can be generated by the 
 shift $N\to N+2m^2$
\begin{align}\label{shift-m}
F_{\rmP}(N,y,m) = F_{\rmP}(N+2m^2,y,0) +   F_{\rm zero} (N,y) \ ,  
\end{align}
where   $ F_{\rm zero}=c_1'+c_2'N +c_3' y$ with the constants $c_i'$ that can be read off  from \re{resum}. 

Another interesting feature of \re{resum} is the appearance of the constant 
 $8\log 2$ in the argument of the  logarithm in \re{resum}. It follows from \re{large-N} that the coefficients of the strong coupling expansion of the free energy involve powers of this constant. At the same time, it is obvious from \re{resum} that all such terms can be eliminated at once by a shift $y\to y-8\log 2$. In terms of the  't Hooft coupling  $\lambda=(4\pi)^2{N/ y}$ this amounts to changing the expansion parameter from $\lambda$ to $\lambda'=(4\pi)^2N/(y+8\log 2)$, or,  equivalently, 
\begin{align}\label{l'}
\lambda' = {\lambda\over 1+{ \log 2 \over 2 \pi^2 N}\lambda}\,.  
\end{align}
This relation (the same as in  \rf{12} and \rf{lapp}) can be interpreted as a finite renormalization of the gauge coupling constant \rf{122}.  
 
Expressed in  terms of the modified coupling constant \re{l'},  Eq.  \re{resum}
reads
\begin{align}\notag\label{resum1}
&F_{\rmP}(N,y,m) =  (N+\tfrac34+2m^2)(N+\tfrac14+2m^2) \log\tfrac{(4\pi)^2 N }{ \lambda'}
\\
 &\qquad\ \ \ \  \ \ \  \ \ \ \ -\log \left[{G(N+\tfrac54+2m^2)G(N+\tfrac74+2m^2)}\right]
 +\tilde c_1+c_2 N+ c_3\,   \tfrac{(4\pi)^2 N }{ \lambda'} \,,
\end{align}
where $\tilde c_1=c_1-   8\log 2  \, c_3$. We verify that the leading term in this expression, $F_{\rmP}= - N^2\log(\lambda') + \dots$, leads to 
$\Delta F=F^{\mathcal N=2}-F^{\mathcal N=4}=N^2\log(\lambda/\lambda') + \dots$, in agreement with \re{hint1}. 
 
Let us  also comment on a possible interpretation of the redefinition of the coupling constant \rf{l'}. It follows from \rf{Zm-int} that, at strong coupling $\l\gg 1$, the dominant contribution to the matrix
 integral comes  from $\tr X^2=2 \sum^N_{n=1} x^2_n = O(\lambda)$ or equivalently $x_n=O(\sqrt\l)$. 
 Examining the interaction potential $S_{\rm int}$ in this limit we find from \rf{Zm}\foot{Here $\log 2$  originates from the finite quantity $
 \eta(1) = \sum_{k=1}^\infty {(-1)^{k-1}\ov k}=\log 2$ that appears in  the expansion  of  (\ref{4.6})
and is effectively   due to the different arguments in the $H$-functions.
 }
\ba
\la{11.2}
S_{\rm int}& = \sum^N_{n=1}\big[4\log H(x_{n} + m) +4\log H(x_{n} - m) -2\log H(2x_{n})\big]\no \\
 &\stackrel{x_{n}\gg 1}{=} 8\log2\,  \sum^N_{n=1} x_{n}^{2}  - (1  + 8 m^2)  \sum^N_{n=1} \log x_{n} + \dots \ .
\ea
Combined  with the Gaussian action $\frac{(4\pi)^{2}N}{\lambda} 
 \sum^N_{n=1} x_{n}^{2} $  in \rf{Zm-int}, the first term on the right-hand side of \rf{11.2} implies a redefinition of the 
gauge coupling  \rf{l'}.  The second term effectively  results  in an extra  order $N$  shift  of the coefficient of the 
$\log \l'$ term in \rf{resum1}:
$(N+\frac34+2m^2)(N+\frac14+2m^2) = N^2 +  N ( 1 + 4 m^2) +O(N^0)$. 
Indeed, the resulting integral over the eigenvalues $x_n$ in \re{Zm-int}  takes the form   
\ba\la{nnz}
Z_N= {1\over N!}\int  \prod_{n=1}^N dx_n \, x_n^2   
 \prod^N_{m>n=1}(x_{n}^{2}-x_{m}^{2})^{2} 
 \,  e^{-\tfrac{(4 \pi)^2 N}{ \l'} \sum_{n=1}^N  x^2_n 
+ (1  + 8 m^2)   \sum^N_{n=1}\log x_{n}+\dots}\,.
\ea
Rescaling the integration variables as 
$x_n\to x_n \sqrt{\l'}$ (with the measure  contributing  the   $(\sqrt{\l'})^{2N^{2}+N}$
factor) we find that the partition function scales at large $\l'$ as
\begin{align}\la{znn}
Z_N \sim e^{-  \big[ N^2 +  N ( 1 + 4 m^2) + \OO(N^0)\big] \log (1/\l' ) } \ , 
\end{align} 
which  is in agreement with \rf{11} and \rf{resum1}.
  
\subsection{Leading nonperturbative correction \la{S2-5}}   

The leading nonperturbative correction to the free energy satisfies the relation \re{246}. Replacing the perturbative function $F_{\rmP}(N,y)$ with its expression \re{resum}, we get from \re{246}
\begin{align}\label{eq-np}
\partial_y^2 F_{\rmNP}(N,y,m) = \Delta_{_N}  F_{\rmNP}(N,y,m) \ 
 { (N+\tfrac34+2m^2)(N+\tfrac14+2m^2)\over (y +  8 \log 2)^2}  \,.
\end{align}
To leading order in $1/N$ the solution to this equation is given by \re{231}
\begin{align}\label{bc-np}
F_{\rmNP}(N,y,m) = N \cosh (2\pi m)\, \FF_{1,\rmNP}(\lambda) + \OO(N^0) \ , 
\end{align}
where $\FF_{1,\rmNP}(\lambda)$ is given by \re{8.29}  with
$\lambda=(4\pi)^2N/y$. We show below that the equation \re{eq-np} supplemented with \re{bc-np} allows us to determine 
subleading corrections to \re{bc-np} and to sum up the series in $1/N$. 
 
Solving \re{eq-np}, it is convenient to introduce an auxiliary function
\begin{align}\label{f-F}
\widehat F(N,y)=F_{\rmNP}(N-\tfrac12-2m^2,\, y-8\log 2,\, m)\ . 
\end{align}
It follows from \re{eq-np} that it satisfies the equation 
\begin{align}\label{eq-f}
\partial_y^2 \widehat F(N,y)=  {N^2-\tfrac1{16}\over y^2}\, \Delta_{_N}  \widehat F(N,y)\,.
\end{align}
Compared to \re{eq-np}, this relation does not involve $m$ and $8\log 2$. In addition, the expression on the right-hand side of \re{eq-f} is even in $N$ and, as a consequence, the  large $N$  expansion of the function $\widehat F(N,y)$ runs in powers of $1/N^2$
\begin{align}\label{f-exp}
\widehat F(N,y) = N \widehat \FF_0(\lambda) + {1\over N}\widehat \FF_1(\lambda) +  {1\over N^3}\widehat \FF_2(\lambda) +\OO\big(\frac{1}{N^5}\big)\,,
\end{align}
where again $\lambda=(4\pi)^2N/y$.
The leading term of the expansion can be obtained by matching \re{f-exp} 
with  \re{f-F} and \re{bc-np}
\begin{align}\label{f0}
\widehat \FF_0(\lambda)=\tfrac{8\sqrt 2}{\pi^{3/2}}\,\l^{-1/4}\,e^{-\sql}\,\cosh(2\pi m) \  
 \Big[1+\OO\big(\tfrac{1}{\sqrt\lambda}\big)\Big] \,.
\end{align}
To find the subleading functions in \re{f-exp}, we substitute \re{f-exp} into \re{eq-f} and compare the coefficients in front of the powers of $1/N$ on both sides. This leads to
\begin{align}\notag \label{ff}
\widehat \FF_1(\lambda) & =\te  \Big(\frac{1}{48} \lambda ^3 \partial_\lambda^3 +\frac{1}{32} \lambda ^2 \partial_\lambda^2   -\frac{1}{16} \lambda  \partial_\lambda\Big)
\widehat \FF_0 (\lambda) \,,
\\[2mm] 
\widehat \FF_2 (\lambda) & =\te   \Big(\frac{1 }{4608}\lambda ^6\partial_\lambda^6 
   +\frac{11 }{7680}\lambda ^5\partial_\lambda^5
+\frac{1}{6144}\lambda ^4\partial_\lambda^4
   -\frac{1}{512} \lambda ^3 \partial_\lambda^3\Big)
\widehat \FF_0 (\lambda) \,,\qquad \dots
\end{align}
As before, the solutions to \re{eq-f} are defined up to zero modes of the operators $\del^2_y$ and $\Delta_{_N}$. 
In distinction to the perturbative part \re{resum}, the zero modes do not contribute to \re{ff}. The reason for this is that $\widehat F(N,y)$ is exponentially small at strong coupling, or, equivalently, $\widehat \FF_k(\lambda) =\OO(e^{-\sqrt\lambda})$ with $k\ge 1$, and the contribution of zero modes is incompatible with this behaviour. 

The relations \re{f0} and \re{ff} allows us to determine the coefficient functions in the large $N$ expansion \re{f-exp}. Applying \re{f-F} we can then derive the  large $N$ expansion of the nonperturbative part of the free energy
(for $N'=N+\tfrac12+2m^2$)
\begin{align}\notag\label{N'}
F_{\rmNP}(N,y,m) &= \widehat F(N+\tfrac12+2m^2,y+8\log 2)
\\[2mm]
&=N' \widehat \FF_0(\lambda'\tfrac{N'}{N}) + {1\over N'}\widehat \FF_1(\lambda'\tfrac{N'}{N})  
 +  {1\over N'^3}\widehat \FF_2(\lambda'\tfrac{N'}{N})  +  \OO\Big(\frac{1}{N'^5}\Big)\ , 
\end{align}
where 
$\lambda'$ was defined in \re{l'}. Here in the second line we took into account \re{f-exp} and replaced
$\lambda=(4\pi)^2N/y$ in \re{f-exp} with $(4\pi)^2(N+\tfrac12+2m^2)/(y+8\log 2)=\lambda'N'/N$. 

Notice that the dependence of $F_{\rmNP}(N,y,m)$ on $m$ enters through the factor of $\cosh(2 \pi m)$ in \re{f0} and 
 $N'=N+\tfrac12+2m^2$. We use \re{f0} to verify  that, for $N\to\infty$ and fixed $\lambda'$, the 
relation \re{N'} reproduces the second, nonperturbative,  term in \re{hint1}.
 
As we will see in a moment, the coefficient functions in \re{N'} scale at strong coupling as $\widehat \FF_k/\widehat \FF_0=\OO(\lambda^{3k/2})$ so that the expansion in \re{N'} is dominated by the  terms of the form $(\lambda^{3/2}/N^{2})^k$. All such terms can be summed up  to all orders to produce a simple expression. 

To show this, we 
notice that the differential operators inside the  brackets in \re{ff} are polynomials in $\lambda \partial_\lambda$. For an arbitrary $k\ge 1$ the expression for $\widehat \FF_k(\lambda)$ looks like
\begin{align}
\widehat \FF_k(\lambda)=   {1\over k!} { (\lambda \partial_\lambda)^{3k}\over 48^k}\,  \widehat \FF_0(\lambda) +\dots\,,
\end{align}
where dots denote terms involving smaller powers of $\lambda \partial_\lambda$. 
Using the explicit form of   $\widehat \FF_0(\lambda)$ in  \re{f0} we find (up to $\OO({1\ov \sqrt\lambda})$ corrections)
\begin{align}
\widehat \FF_k(\lambda) =  {1\over k!} \Big(-{\lambda^{3/2}\over 384}\Big)^k\, \widehat \FF_0(\lambda)  \,.
\end{align}
Together with \re{f-exp} this leads to
\begin{align}\label{F-np-resummed} 
F_{\rmNP}(N,y,m)   
&= \tfrac{8\sqrt 2}{\pi^{3/2}}\,N \cosh(2\pi m)  \,({\l'})^{-1/4} \, e^{-\sqrt{\l'}-{{\lambda'}^{3/2}\over 384 N^2}} + \dots\,.  \end{align}
To be precise,  this relation holds for $N$ and $\lambda'$ going to infinity with $(\lambda')^{3/2}/N^2$ kept  fixed  (cf. footnote \ref{foot7}).
In this limit, $N'$ approaches $N$  and \re{N'} coincides with \re{f-exp} with $\lambda$ replaced by $\lambda'$. One can go beyond this approximation and systematically include subleading corrections to \re{F-np-resummed}.

\section{Circular Wilson loop \la{S3}}

In this section, we apply localization to compute 
the expectation value of a  circular BPS  Wilson loop in the mass-deformed $\mathcal N=2$ $Sp(2N)$ theory. 
It is given by a matrix integral that is similar to \re{Zm-int}
\begin{align}\label{W_N}
W_N = {1\over Z_N} \int  \mc DX\,  \tr e^X \,   e^{-\frac{8\pi^{2}}{g_{{{\rm YM}}}^{2}}\tr X^{2}-S_{\rm int}(X,m)} \,, 
\end{align}
where $Z_N$ is the partition function \re{Zm-int} and interaction action is given by \re{Sint}.  Compared  to 
 \re{Zm-int}, the integral in \re{W_N} contains the  extra  factor of $\tr{e^X}$.

 We have seen in the previous section that the use of the Toda equation simplifies significantly the derivation of the  large $N$ expansion of the free energy. As we  find below,  the same is true  for the circular Wilson loop. We show that \re{W_N} satisfies a non-trivial finite-difference equation which allows us to calculate $W_N$ as an expansion in  large $N$. 

To derive this equation, it is convenient to generalize \re{W_N} by introducing an infinite set of parameters $\boldsymbol t=(t_2,t_3,\dots)$  to define 
\begin{align} \la{ttt}
W_N(\boldsymbol t) 
&={1\over Z_N(\boldsymbol t)} \int \mc DX \, \tr e^{ X}  \, e^{-\frac12 \sum_{n\ge 1} t_{2n} \tr (X^{2n})}\ . 
\end{align} 
Here $Z_N(\boldsymbol t)$ is given by the same matrix  integral without  the  $\tr e^{ X}$ factor.  The  expression  \rf{ttt} coincides with \re{W_N} after one identifies the parameters $t_{2n}$ with the coefficients in front of $\tr(X^{2n})$
terms 
 in the exponent of \re{W_N}, e.g.,  $t_2=y=(4\pi)^2/\gym^2$, etc.
 
 The reason  for introducing the parameters $\boldsymbol t$ is that, upon expanding $\tr e^X $ into traces of (even) powers of $X$, the function $W_N(\boldsymbol t)$ can be obtained from a logarithm of the partition function by applying the following 
  linear differential operator
\begin{align}\label{W-Z} 
W_N(\boldsymbol t) &=2N - 2\sum^\infty_{k= 1} {1\over (2k)!}{\partial \over \partial t_{2k}}\log Z_N(\boldsymbol t)\ . 
\end{align}
For $t_2=y$ and arbitrary $t_{2k}$ (with $k>1$) the generalized partition function $Z_N(\boldsymbol t)$   satisfies the Toda equation \re{Toda}.

Applying the differential operator in  
\re{W-Z} to both sides of \re{Toda} and setting  the parameters $\boldsymbol t$ equal  to their values corresponding to \re{W_N}, we obtain the following  finite-difference equation for $W_N$ in  
 \rf{W_N} 
\begin{align}\notag\label{W-eq}
\partial_{y}^2 \, W_N &= \left( W_{N+1}-2 W_{N}+ W_{N-1} \right){Z_{N+1} Z_{N-1} \over Z_N^2 }
\\[2mm]
&= - \left( W_{N+1}-2 W_{N}+ W_{N-1} \right)\partial_{y}^2F_N\,.
\end{align}
Here in the second relation we applied \re{Toda} and used that  $\log Z_N=-F_N$.
Notice that the relation \re{W-eq} is insensitive to the form of the interaction action \re{Sint} and, therefore, holds, in particular,   
 in both  $\mathcal N=4$ and $\mathcal N=2$ theories.

The relation \re{W-eq} is remarkably similar to \re{246}. The important difference between the two equations is that   \re{W-eq} is exact and  holds for an arbitrary coupling. In addition, solving \re{246} we looked for a solution (nonperturbative part of the free energy) that scales at strong coupling as $\OO(e^{-\sqrt\lambda})$. This boundary condition does not apply to \re{W-eq}.

\subsection{Toda equation in $\mathcal N=4$ $Sp(2N)$ theory \la{S3-1}}

Let us examine the Toda equation \re{W-eq} for the circular Wilson loop in $\mathcal N=4$ theory. 
Replacing the free energy with its expression \re{F-N=4} we get from \re{W-eq}
\begin{align} \label{eq-0}
\partial_{y}^2 \, W_N&=  ( W_{N+1}-2 W_{N} + W_{N-1})\, {(N+\tfrac14)^2-\tfrac1{16}\over y^2}\,.
\end{align}
We look for a general solution to this equation in the form of   a $1/N$ expansion
\begin{align}\label{W-large-N}
W_N^{\mathcal N=4}(\lambda)= N W^{(0)}(\lambda) +W^{(1)}(\lambda) + {1\over N} W^{(2)}(\lambda) + \OO\big({1\ov N^2}\big)\,.
\end{align}
At zero coupling we have $W_N(0)=\tr \mathbf 1= 2N$, or, equivalently, $W^{(0)}(0)=2$ and $W^{(k)}(0)=0$ for $k\ge 1$. 

Substituting \re{W-large-N} into \re{eq-0} and matching the coefficients of $1/N$ on both sides we obtain a system of differential equations for 
the functions $W^{(k)}(\lambda)$. Supplemented with the boundary conditions at zero coupling, their solutions are
\begin{align}  \notag
\label{W-W0}
& W^{(1)}(\lambda) =  \tfrac14 (1+\lambda\partial_\lambda) W^{(0)}(\lambda) - \tfrac12\,, \qquad 
 \\[2mm]
 & 
W^{(2)}(\lambda) =  \Big({\tfrac{1}{16} \lambda ^2\partial_\lambda^2+\tfrac{1}{48} \lambda ^3 \partial_\lambda^3}\Big)
   W^{(0)}(\lambda )\,,  \qquad 
   \dots\ . 
\end{align}
These relations are analogous to those for the free energy \re{88}. 

In the $\mathcal N=4$ SYM  theory with the $Sp(2N)$ gauge group, 
the leading term of the large $N$ expansion \re{eq-0} is  well-known   \ci{Fiol:2014fla,Giombi:2020kvo}\foot{In our notation where 
the  $N$-factor is extracted (cf. \rf{W-large-N})  this  is effectively  the same as in  the  $SU(2N)$   case \ci{Erickson:2000af,Drukker:2000rr}.}
\begin{align}\label{W0} 
W^{(0)}(\lambda) = \te {4\over\sqrt\lambda} I_1(\sqrt\lambda)\,.
\end{align}
Substituting this expression into \re{W-W0} we get
\begin{align}\label{W00} 
W^{(1)}(\lambda) =\tfrac12\big[{I_0(\sqrt\lambda)-1}\big]\,,\qquad
W^{(2)}(\lambda) =\tfrac{1}{96} \l\,I_2(\sqrt\lambda) \,,\qquad \dots\ . 
\end{align}
The large $N$ expansion \re{W-large-N} can be resumed to all orders in $1/N$ to yield the following integral representation
\begin{align}\label{W0-int}
W_N^{\mathcal N=4}= {8N\over\lambda} \oint {dz \over 2\pi i}\,  e^{-z+{\lambda\over 16N}}\, 
{1-{\lambda \over 8Nz}\over 1-{\lambda \over 16Nz}}\Big[ 
1-\Big({1-{\lambda\over 8Nz}}\Big)^{2N}\Big]\,,
\end{align}
where the integration contour encircles the origin in anti-clockwise direction.
Using that  $\lambda=(4\pi)^2 N/y$, it is possible to show that this expression verifies the Toda equation \re{eq-0}. 

Changing the integration variable in \re{W0-int} as $z\to 8Nz/\lambda$ we get from \re{W0-int}
\begin{align}\label{Lag}
W_N^{\mathcal N=4} &=2 e^{{\lambda\over 16N}} \sum_{i=0}^{N-1}\oint {dz \over 2\pi i}  \ e^{-{\lambda\over 8N}z}\ z^{-2-2i}(z-1)^{1+2i}
=2 e^{{\lambda\over 16N}} \sum_{i=0}^{N-1} L_{2i+1}\Big({-{\lambda\over 8N}}\Big)\,,
\end{align}
where $L_n(x)$ is the Laguerre polynomial. This relation is exact and 
holds in the  $\mathcal N=4$ $Sp(2N)$ theory for an arbitrary $N$ and  $\lambda$. 

At strong coupling, the integral in \re{W0-int} can be evaluated using a  saddle point  approximation.
 A close examination shows that there are two saddle points  
$z_\pm=\frac{1}{2} \sqrt{\lambda } \big(\frac{\sqrt{\lambda }}{8 N} \mp \sqrt{\frac{\lambda }{64
N^2}+1}\big)$. The integration in the vicinity of $z=z_+$ yields the contribution that scales as $\OO(e^{\sqrt\lambda})$ whereas the contribution of
$z=z_-$ behaves as $\OO(e^{-\sqrt\lambda})$.
 In the double scaling limit  
  $N\to \infty$ and $\lambda\to \infty$ with 
$\lambda^{3/2}/N^2$ held fixed, we find that  (cf. footnote \ref{foot7})
\begin{align}\label{W-N4-str}
W_N^{\mathcal N=4} = N\sqrt{\tfrac{8}{ \pi}} \, \lambda^{-3/4} e^{\sqrt\lambda+{\lambda^{3/2}\over 384N^2}} +\OO\Big(e^{-\sqrt\lambda -{\lambda^{3/2}\over 384N^2}}\Big)\,.
\end{align}
Here the second term can be formally obtained from the first one by replacing $\sqrt\lambda\to-\sqrt\lambda$.
Following the  terminology adopted for the free energy \re{npp}, the two terms on the right-hand side of \re{W-N4-str} 
may be interpreted as representing the  perturbative and nonperturbative contributions to the Wilson loop. 

In contrast to the free energy \re{large-N}, the perturbative contribution
 in \re{W-N4-str} scales as $\OO(e^{\sqrt\lambda})$ (rather than $\OO(\log\lambda)$). At the same time, it is interesting to notice that nonperturbative corrections to \re{W-N4-str} and \re{F-np-resummed} involve the same exponentially small factor
 (but here it has  an imaginary prefactor, cf. \ci{Drukker:2006ga,Zarembo:2016bbk,Beccaria:2021ism}). 

\subsection{Difference of  Wilson loops in $\N=2$ and $\N=4$  theories \la{S3-2}}

As we have demonstrated in section~\ref{S2-3}, 
 the Toda equation for the partition function \re{Toda} is powerful enough to predict the subleading terms in the $1/N$ expansion of the difference free energy \re{Delta-F} in terms of the leading one $\FF_1(\lambda, m)$  (see Eqs.~\re{88}). We can repeat the same analysis for \re{W-eq} to show that similar relations also hold for the  coefficients in  the $1/N$ expansion of the Wilson loop. 

In a close analogy with the difference free energy function \re{2v}, we define the difference between the circular Wilson loops in
the  $\mathcal N=4$ and $\mathcal N=2$ models
\begin{align}\label{DW-exp}\notag
\Delta W  &= W_N^{\mathcal N=2} - W_N^{\mathcal N=4} 
\\ &
=  \Delta W^{(1)}(\lambda,m) + {1\over N}  \Delta W^{(2)}(\lambda,m)+ {1\over N^2}  \Delta W^{(3)}(\lambda,m) + \OO\big({1\ov N^3}\big)\,.
\end{align}
Here we took into account that the leading $\OO(N)$ term  in the difference cancels out 
 due to the planar equivalence of the two theories. 
 Notice that $\Delta W_N$ vanishes at zero coupling, i.e. for $\lambda=0$.

Substituting $W_N\to  W_N^{\mathcal N=2}=   W_N^{\mathcal N=4}+ \Delta W $ and $F_N\to  F_N^{\mathcal N=2}=F_N^{\mathcal N=4}+ \Delta F $ into \re{W-eq},  using that   $\lambda=(4\pi)^2N/y$ and comparing  the coefficients of powers of $1/N$ on both sides of \re{W-eq} we  get a system of linear equations for the functions $\Delta W^{(k)}(\lambda,m)$ with $k\ge 1$. These equations involve the functions $\FF_k$ and $W^{(k)}$ which enter the large $N$ expansions of $\Delta F$ and $W_N^{\mathcal N=4}$, respectively (see Eqs.~\re{Delta-F} and \re{W-large-N}). According to \re{88} and \re{W-W0}, they, in turn, can be expressed in terms of the leading functions $\FF_1(\l,m)$ and $W^{(0)}(\l)$.

Combining these relations and supplementing them with the boundary condition at zero coupling, $ \Delta W^{(k)}(\lambda=0,m) =0$, we get after some algebra
\begin{align}\notag
\la{7.23}
 \Delta W^{(1)} (\lambda,m) &= -\tfrac{1}{2}\int_0^\l d\l \, \sqrt\l\,  I_1(\sqrt\lambda)\,\GG(\l,m)\,, 
\\
\Delta W^{(2)}(\lambda,m) &= -\tfrac{1}{8}\, \l^{3/2} I_1(\sqrt\lambda)\,\Big[\GG(\l,m)-\l\,\GG^2(\l,m)\Big]\,, 
\end{align}
where  $\GG(\l,m)$ was defined in \re{6.12} and \re{8.20}. 
Here we replaced $W^{(0)}(\l)$ with its expression \re{W0}.
Remarkably, higher terms of the $1/N$ expansion in \re{DW-exp} admit a closed form representation in terms of the functions $W^{(0)}(\l)$ and $\GG(\l,m)$, though  the corresponding expressions are lengthy, e.g.,
\ba \notag
& \Delta W^{(3)}(\lambda,m) =   -\tfrac{1}{192} \lambda ^3 I_0(\sqrt{\lambda }) \left(-3 \GG^2+2 
\lambda\,  \GG^3+\GG {'}\right)
\\[2mm]
&\ \ \ \ \   -\tfrac{1}{384} \lambda 
^{5/2} I_1(\sqrt{\lambda }) \Big(\GG-12 \GG^2+16 
\lambda \GG^3   +4 \GG {'}-24 \lambda  \GG \GG {'} 
+24 \lambda ^2 \GG ^2 \GG {'}+4 
\lambda  \GG {''}\Big) , \la{7.31}
\ea
where  primes denote again derivatives with respect to $\lambda$.
The relations \re{7.23} and \re{7.31}   can be considered as the counter-parts of the analogous relations \re{6.12} and \re{6.14} for the free energy. 
Being combined together with \re{6.12} and \re{8.20}, they allow us to compute the Wilson loop in the $\mathcal N=2$ model for an arbitrary coupling. 

The relations \re{7.23}, \re{7.31},  etc.,   can be used  
to derive the strong coupling expansion of the Wilson loop to any given order in $1/N$. They are not suitable, however,  for discussing a  resummation of  the large $N$ expansion. In the next subsection we apply \re{W-eq} to compute the circular Wilson loop in $\mathcal N=2$ $Sp(2N)$ theory at strong coupling. 

\subsection{Toda equation  in $\mathcal N=2$ $Sp(2N)$ theory  \la{S3-3}}

According to \re{npp}, the free energy in $\mathcal N=2$ $Sp(2N)$ theory is given at strong coupling by the sum of perturbative and nonperturbative pieces.  
Correspondingly, substituting \re{npp} into the Toda equation \re{W-eq} we can look for its solution in the form
\begin{align}\label{W-n+np}
W_N^{\mathcal N=2}= W_{N,\rmP} + W_{N, \rmNP} \,,
\end{align}
where the second (nonperturbative) term is  exponentially small as compared to the first (perturbative) term. As we will see in a moment, 
the two terms on the right-hand side  
 account  for the corrections $\OO(e^{\sqrt\lambda})$ and $\OO(\l^0)$, respectively. 

It follows from \re{W-eq} and \re{resum} that the perturbative contribution $W_{N,\rmP}$ satisfies
\begin{align}\label{eq-W-P} 
\partial_{y}^2  W_{N,\rmP} &=  \lr{W_{N+1,\rmP} -2 W_{N,\rmP} + W_{N-1,\rmP}  }\, {(N+\tfrac12+2m^2)^2-\tfrac1{16}\over (y+8\log 2)^2}\,.
\end{align}
As in the case of the free energy \re{Toda-P}, it is assumed here that $W_{N,\rmP}$ is a function of $y$ (rather than $\lambda=(4\pi)^2N/y$).
Let us compare this relation with \re{eq-0}. We observe that the two equations coincide after one applies the shift $N\to N+\tfrac14+2m^2$ and $y\to y+8\log 2$ to \re{eq-0}. This implies that, up to the contribution of the zero modes, the solutions to \re{eq-W-P} and \re{eq-0} are related to each other through the same transformation  
\begin{align}\label{rel-W}
W_{N, \rmP} &=W^{\mathcal N=4}_{N+\frac14+2m^2}(y+8\log 2)\ .
\end{align}
 It is important to emphasize that this relation only holds at strong coupling up to exponentially small (nonperturbative) corrections. 
 The relation \re{rel-W} is analogous to \re{shift-m} in that the dependence of the Wilson loop on the mass parameter  can be generated by the shift $N\to N+2m^2$. Since  the  perturbative part of $W_N^{\mathcal N=4}$  scales  
 as $\OO(e^{\sqrt\lambda})$, the contribution of the  zero modes to \re{rel-W} is exponentially small. 

Viewed as a  function of $N$ and $\lambda={(4\pi)^2 N /y}$,  \ $W_{N, \rmP}$ in \re{rel-W} takes the form similar to \re{N'} 
\begin{align}\label{Wp-shift}
W_{N, \rmP}(\lambda) &=W^{\mathcal N=4}_{N'}(\lambda'\tfrac{N'}{N}) \,,
\end{align}
where $N'=N+\tfrac12+2m^2$ and $\lambda'$ is defined in \re{l'}. In the double-scaling limit \re{W-N4-str}, $W_{N,\rmP}$ coincides with  \re{W-N4-str}. 
 
\subsection{Leading nonperturbative correction  \la{S3-4}}

According to \re{W-N4-str}, the 
leading nonperturbative correction to the Wilson loop in the $\mathcal N=4$ theory is suppressed by the factor of 
$e^{-2\sqrt\lambda}$ as compared to the perturbative contribution. 
Interestingly,   as we  show below,  in the $\mathcal N=2$ model the situation is different in that 
the leading nonperturbative correction to the difference  $\Delta W$ 
 scales as  $\OO(\lambda)$, i.e. 
   $W_{N,\rmNP}=\OO(\lambda)$  (compared to 
   $\OO(e^{-\sqrt\l})$  in  the $\mathcal N=4$ case).
   This correction is still exponentially suppressed relative  to the perturbative one $\sim e^{\sql}$. 

To demonstrate this, we apply \re{7.23} and replace the function $\GG(\l,m)$ with its expression \re{calG}. At strong coupling the Bessel function in \re{7.23} is given by the sum of two terms that behave as $e^{\sqrt\l}$ and $e^{-\sqrt\l}$. As a consequence, the leading nonperturbative correction to \re{7.23} comes from the interference of the former term and the nonperturbative $\OO(e^{-\sqrt\l})$ correction to $\GG(\l,m)$.
Taking into account \re{calG} we get from \re{7.23}
\begin{align}
 \partial_\l \Delta W_{\rm np}^{(1)} (\lambda,m) = -\tfrac{1}{2} \sqrt\l\,  I_1(\sqrt\lambda)\,\GG_{\rm np}(\l,m) = - \tfrac{1}{ \pi^2}\cosh(2\pi m) + \OO\big(\tfrac{1}{\sqrt\lambda}\big)\,,
\end{align}
wherefrom $\Delta W_{\rm np}^{(1)} (\lambda,m)=\OO(\lambda)$.
In a similar manner, it follows from the second relation in \re{7.23} that 
\begin{align} \label{DW-np}
& \Delta W_{\rm np}^{(2)}(\lambda,m) = -\tfrac{1}{8}\, \l^{3/2} I_1(\sqrt\lambda)\,\Big[\GG_{\rm np}(\l,m)-2\l\,\GG_{\rm np}(\l,m) \GG_{\rm p}(\l,m) \Big]+\dots\,, 
\end{align}
where dots denote subleading corrections. 

Replacing $\GG_{\rm p}(\l,m)$ in \re{DW-np} with its expression given by the first term in  \re{calG} 
we notice that $\Delta W_{\rm np}^{(2)}(\lambda,m)$ is proportional to $\partial_\l \Delta W_{\rm np}^{(1)} (\lambda,m)$ 
\begin{align}
\Delta W_{\rm np}^{(2)}(\lambda,m) =  \Big( \tfrac12 +2m^2 -  \tfrac{\log 2}{ 2\pi^2}  \l\Big)  \l\partial_\l \Delta W_{\rm np}^{(1)} (\lambda,m)\,.
\end{align}
This leads to
\begin{align}\label{assum}
\Delta W_{\rm np} =
\Delta W_{\rm np}^{(1)}+ {1\over N} \Delta W_{\rm np}^{(2)} + \OO\big({1\ov N^2}\big)=
\Delta W_{\rm np}^{(1)}\Big(\lambda'\big(1+\tfrac{\frac{1}{ 2} + 2m^2 }{ N}\big)\Big)+ \dots\ , 
\end{align}
where $\l'$ is as  in \re{l'}. Although this relation holds at order $\OO(1/N^2)$, we assumed that higher order corrections in $1/N$ only modify the argument of the leading term. 
 
To justify \re{assum} we apply the Toda equation \re{W-eq}. Substituting
 \re{W-n+np} and taking into account \re{eq-W-P} we get from \re{W-eq}
\begin{align}\notag\label{W-np-eq}
 \partial_{y}^2 W_{N,\rm np} =&{} \lr{ W_{N+1,\rm np} -2 W_{N,\rm np} + W_{N-1,\rm np} }\, {(N+\tfrac12+2m^2)^2-\tfrac1{16}\over (y+8\log 2)^2} 
 \\
&{} - \lr{ W_{N+1,\rm p}-2 W_{N,\rm p}+ W_{N-1,\rm p} }\partial_{y}^2  F_{N,\rm np} \,,
\end{align} 
where $W_{N,\rm p}(y,m)$ and $ F_{N,\rm np}(N,y,m)$ are given by \re{rel-W} and \re{F-np-resummed}, respectively. Here we neglected the subleading terms proportional to the product of $F_{N,\rm np}$ and $W_{N,\rm np}$. 

As before, we can simplify the relation \re{W-np-eq} by applying the shifts $N\to N-{1\ov 2} -2m^2$   and 
$y\to y-8\log 2$.  Introducing the 
function $w_N(y)$ defined by 
\begin{align}\label{W-w}
 W_{N ,\rm np} (y)=w_{N+\frac12+2m^2}(y+8\log 2) \,,
\end{align}
we find from \re{W-np-eq}, \re{rel-W} and \re{f-F} that $w_N(y)$ satisfies
\begin{align}\label{w-eq}
 \partial_{y}^2 w_{N} &{}= \te \lr{ w_{N+1} -2 w_{N} + w_{N-1} }{ N^2-\tfrac1{16}\over y^2} 
 - \lr{ W^{\mathcal N=4}_{N+1/2}-2 W^{\mathcal N=4}_{N- 1/2}+ W^{\mathcal N=4}_{N-3/2} }\partial_{y}^2  \widehat F_{N} \,.
\end{align}
At strong coupling,  $W^{\mathcal N=4}_{N}$ is given by \re{W-N4-str} whereas the function $\widehat F_{N}$ can be found from \re{F-np-resummed} by replacing $\lambda'$ with $\lambda$. Using    the large $N$ expansion,
 $w_N=w^{(1)}(\lambda) +{1\ov N} w^{(2)}(\lambda) + \dots$,  we obtain from \re{w-eq} that the leading term $w^{(1)}(\lambda)$
 coincides with $\Delta W_{\rm np}^{(1)} (\lambda,m) $ and  reads
\begin{align}\label{w1-np}
w^{(1)}(\lambda) = \Delta W_{\rm np}^{(1)} (\lambda,m) =- \tfrac{1}{ \pi^2}\l\, \cosh(2\pi m) + \OO(\sqrt\lambda)\,. 
\end{align}
Combining this relation together with \re{W-w} we arrive at \re{assum}. We verified that the relations \re{assum} and \re{w1-np} correctly reproduce the leading corrections to $\Delta W_{N,\rm np}(\l,m)$ of the form $\lambda^{k+1}/N^k$ and $m^2 \lambda^{k}/N^k$ for $k\ge 0$.
The functions $w^{(k)}$ (with $k\ge 2$) give subleading corrections to $\Delta W_{N,\rm np}(\l,m)$ that are suppressed by powers of $1/\sqrt\lambda$ as compared to those coming from \re{w1-np}.  

 \section{Dual  string theory  interpretation \la{S6}}
  
Let us now  comment on    the  dual 
string theory  interpretation of the  strong-coupling expansions
derived on the gauge theory side. 
The  string  theory  for the $Sp({2N})$  FA-orientifold  theory (i.e. the  $\N=2$  $Sp({2N})$  superconformal model 
with 4 fundamental and 1  antisymmetric hypers)  can be defined 
   \ci{Fayyazuddin:1998fb,Aharony:1998xz,Ennes:2000fu}  
  using    a near-horizon  limit of the  system of 
     $2N$ D3-branes  with  8 D7-branes  stuck on one 
O7-plane. The effective presence of   D7-branes   introduces the  new D3-D7 open string sector (with massless   modes  being related to the fundamental hypermultiplets in the corresponding gauge theory).   Equivalently, it can be defined 
as 
 the type IIB   superstring on  the orientifold 
 AdS$_5{\times}  S'^5$ where $S'^5=S^5/\Z_{2,\ori}$.\foot{The orientifold group  is  defined as 
$\Z_{2,\ori}=\{1,\, I_{45}\, \Omega\,  (-1)^{F_L}\}$.  It contains  the  $\Z_2$  orbifold action:
the  inversion $I_{45}$ 
acts on the 2-plane of  $\mathbb R^6$  (with directions 4,...,9) transverse to the D3-branes  as
$\  x_{4,5} {\rightarrow}- x_{4,5}$.
The fixed-point set of this $\Z_{2}$ is the hyperplane $x_{4,5}=0$,
which corresponds to the position of  the   8   D7-branes and O7-plane.
 In the   near-horizon limit  the 
  $\Z_{2}$ orbifold part of $\Z_{2,\ori}$ acts   on the coordinates of $S^5$
(with the metric
$     ds_5^2  =d\theta_1^2 +  \sin^2\theta_1\, d\varphi_1^2  + 
 \cos^2\theta_1 \,dS_3 , \ \   dS_3= d\theta_2^2 + \cos^2\theta_2\, d\varphi_2^2+\sin^2\theta_2\,d\varphi_2^2
$)   as   $\varphi_1 \to \varphi_1 + \pi$. Then $\theta_1=0$  subspace   is the
collection of  conical singularities represented by $S^3$. 
}
One may interpret the resulting   theory as  containing 
D7  branes    wrapped on AdS$_5{\times} S^3$  where $S^3$ is fixed-point locus  of $\Z_{2,\ori}$.

The dual string perturbation theory  will then  involve  both 
 closed-string  and open-string  world-sheet topologies, i.e. corrections 
 of both even and odd    powers in $\gs$, corresponding to  even  and odd powers of $1/N$ on the   gauge theory side. The orientifolding implies the 
 presence  of  contributions of   non-orientable  surfaces with  crosscups.\foot{
 Note that in   the  $Sp(2N)$  $\N=4$ SYM case  dual to   string in 
 AdS$_5{\times}  \mathbb{RP}^5$ \cite{Witten:1998xy}
  all odd-power  $1/N$  contributions   should come from crosscups 
  while in the  $Sp(2N)$  $\N=2$ \FA   model 
  there should  be also $1/N^{2k+1}$ contributions  from world sheets  with  boundaries
    reflecting  the presence of D7-branes   (cf. 
       \ci{Aharony:1999ti}).}

 Accounting for  the open string  sector term 
  in the  dual string theory effective action (that here 
  may be  interpreted  as the  D7+O7-brane world-volume action)
   allowed   \ci{Aharony:1999rz,Blau:1999vz} to give 
  a  holographic  interpretation of the order $N$ term in  the  (super)conformal anomalies of the $Sp(2N)$  \FA theory. 

To recall, in  the  $\N=4$   and $\N=2$   $Sp(2N)$ 
theories  
the a  and c conformal anomaly coefficients  are given by\foot{To recall, 
the  conformal anomaly relation is 
$(4\pi)^{2} \vev{T\indices{^{m}_{m}}} = -\text{a}\,E_{4}+\text{c}\,C^{2} = 
 (\cc-\aa) R_{mnkl}^{2} 
+ \OO(R^2_{mn}, R^2)
$,  where 
$C^{2}=R_{mnkl}^{2}-2R_{mn}^{2}+\frac{1}{3}R^{2} $, \ \ 
$E_{4}=R_{mnkl}^{2}-4R_{mn}^{2}+R^{2}$.
In the $\N=2$ superconformal models  with gauge group $G$  one has 
 $4(2\text{a}-\text{c}) = \dim G$ \cite{Shapere:2008zf}.
In particular,  in the present  case  $\dim [ Sp(2N) ]= N(2N+1)$.
To capture the $\cc-\aa$  combination it is  sufficient to assume that  the boundary metric is 
Ricci flat,  $R_{mn}=0$. 
}
\ba \la{41}
&\N=4 \ \text{SYM}: \qquad  \te  \aa=\cc =  \frac{1}{2}N^{2}+\frac{1}{4}N =\frac{1}{2} ( N + \four)^2 - {1\ov 32} \ ,   \\
&\N=2 \  \text{FA}:  \qquad \ \ \ \te  \aa= 
\frac{1}{2}N^{2}+\frac{1}{2}N-\frac{1}{24}  \ , \qquad \ \ \cc= \frac{1}{2}N^{2}+\frac{3}{4}N-\frac{1}{12} \la{42}  \ . 
\ea
On the superconformal 
gauge theory side  the a-anomaly coefficient   appears in  the UV   divergent 
part  free energy on $S^4$  as 
\be \la{411}
F= 4 \aa \log ( \L_{\rm UV} \rr) +  ...  \ee
where $\L_{\rm UV}$ is a  UV cutoff and $\rr$ is the radius of  $S^4$  and dots stand  for possible finite contributions. 
In the   localization matrix model   computation of $F$  the UV divergence is automatically subtracted. In the $\N=4$ SYM  case  described by  Gaussian matrix model  the dependence on the radius $\rr$  is  correlated  with dependence on $\l$ (they enter the free part  of the  
action in \rf{4.1} in combination $\rr^2\ov \l$). As a result, 
\be \la{412}
F= -2  \aa \log ( \l  \rr^{-2}) +... \ ,  \ee
where  as in \rf{411}  the  anomaly 
 coefficient $\aa$   controls  the   dependence on $\rr$.
While the  dependences on $\l$   and on  the $S^4$ radius $\rr$ 
 are a priori correlated only in the $\N=4$ SYM or Gaussian matrix model case (see 
Appendix A of \cite{Beccaria:2021ism})  it turns out that in the $Sp(2N)$  FA model 
this applies also to the subleading order $N$ term:  the $N^2 + N$   combination in $\aa-$anomaly in \rf{42} is the same as  in the  coefficient of the $\log \l$ term in \rf{11}.

In general, the 
gauge theory  free  energy on $S^4$   should  be reproduced by the string   partition $Z_{\rm str}$
 function on AdS$_5{\times}  S'^5$  where $S^4$ is the boundary of  AdS$_5$.
     Since this is a  homogeneous space, the 
  field  theory intuition  suggests  that  the result   should  be proportional to the volume of AdS$_5$
    space.  
 The latter is  IR  divergent,\footnote{The regularized volume  of odd-dimensional AdS space is
 $\text{vol}(AdS_{2n+1}) = \frac{2(-\pi)^{n}}{\Gamma(n+1)}\,\log (\Lambda_{\rm IR}  \rr)$ where $\rr$ is the radius of the boundary sphere  and $\L_{\rm IR}$ is an IR cutoff.}
 $\Vol({\rm AdS}_5) = \pi^2 \log (\L _{\rm IR} \rr)$.
 In  \ci{Russo:2012ay} it was suggested  to use a particular string tension related 
  IR cutoff   $\L _{\rm IR}  \sim {1\ov \sql}$  so that\foot{Here  we formally ignore  mismatch of dimensions to indicate that the dependence on $\l$ is correlated  with that on $r$.}
 \be \la{111}
\Vol({\rm AdS}_5) = \pi^2 \log (\L _{\rm IR} \rr)\ \  \to \ \  \te - \ha \pi^2 \log (\l \rr^{-2}) \ .
\ee 
The  leading 2-sphere topology   contribution to   $Z_{\rm str}$  
   may be represented as  the type IIB supergravity action (plus $\a'$-corrections).  
Starting   with 10d supergravity action  evaluated on \adss  one   reproduces the 
leading $N^2$  term in the  localization result  for $SU(N)$  $\N=4$ SYM free energy, 
\be  \la{222}
SU(N): \ \ \ \ \   F_{\N=4}= - \tfrac{1}{2}\,(N^2-1) \log ({ \l \rr^{-2}})  + C(N)  \ . \ee
To recall, compactifying 10d   supergravity action on $S^5$ (of radius $L)$  one finds 
(after accounting for a 5-form dependent boundary term \ci{Kurlyand:2022vzv})
   the  familiar  5d  action ($\Vol(S^5) = \pi^3$)
   \be \la{555}
   S_{10}= -\tfrac{\Vol(S^5)}{2\kappa_{10}^{2}} \int d^{5}x\,\sqrt{-g}(R_5+12 L^{-2})
\ , \quad 2 \kappa^2_{10} = ( 2 \pi)^7 \alpha'^4  g_s^2\ , \quad 
L^{4}=4\pi g_{s} N \alpha'^{2} \ . \ee
Using that   for AdS$_5$  one has $R_5=-20 L^{-2}$   and \rf{111}
we get 
\be\la{32}
 S_{10}=  \tfrac{1}{\pi^{2}}\,  N^2\,  \text{Vol}({\rm AdS}_{5})    
 = N^{2}\,\log(\L_{\rm IR} \rr)  \ ,  \ee
 that then matches  the $N^2$ term in \rf{222}.
 
 One can  also reproduce the subleading $(-1)$  term  in the a-anomaly coefficient in \rf{222}
 by accounting for 1-loop (torus)  contribution to string partition function which in the  maximally supersymmetric \adss  case   happens to be given  
   by the sum of the 1-loop  contributions of just the 10d supergravity modes \ci{Beccaria:2014xda} (see also \ci{Mansfield:2002pa}). 
 
 Below we will first present the analogous dual string  computation of $F$ in the $\N=4$ $Sp(2N)$   SYM   theory  and then discuss the $\N=2$ $Sp(2N)$  FA   theory.
 
 \subsection{Free energy    in $\N=4$    theory dual to  IIB string on AdS$_5{\times}  \mathbb{RP}^5$ \la{S6-1}}

The  $Sp(2N)$  SYM theory   can be realised on $2N$  D3-branes  and O3 plane 
and is dual to the IIB string on an  orientifold  of AdS$_5\times S^{5}$ or 
 AdS$_5{\times}  \mathbb{RP}^5$ \cite{Witten:1998xy}.
The dual string theory explanation of  the expression for  its free energy 
\rf{4} or the  conformal a-anomaly \rf{41}  is  as follows. 
The $N^2$ term  comes from  the classical 10d supergravity action  just 
as in  the $SU(N)$  case  above. 

The shift $N \to N + \four$ in \rf{41}  explaining  the order $O(N)$ term in the  SYM conformal anomaly 
 in  \rf{41}   may be attributed to the  redefinition of the 
  D3 brane charge due to the presence  of  the O3 planes \ci{Blau:1999vz}:   O3-planes carry
fractional RR charge $\frac{1}{4}$ \cite{Polchinski:1996na,Mukhi:1997zy}. 
Equivalently, from the flat-space perspective,  this shift  may be interpreted as   being due to   crosscup contributions. 
 In  view of this shift the AdS  radius  (given by  $
L^{4} = 4\pi g_{s}\alpha'^{2} N 
$ in the $SU(N)$ case) is   now  identified as 
\cite{Blau:1999vz,Giombi:2020kvo} 
\be \la{1xx}
L^{4} = 4\pi g_{s}\alpha'^{2}(2N+\tfrac{1}{2}) \ . 
\ee
This  leads to 
 $(N+ \four)^2$ as the  coefficient of the AdS$_5$ volume  in the on-shell value of the 10d supergravity action  and as a result we match the $N^2+\ha N$ terms in the free energy in \rf{4}. 
 
Below   we will  provide the explanation for  the remaining  $ (- {1\ov 32})$ 
term in a$-$anomaly coefficient  in \rf{41}   as originating from the   1-loop contributions of the short 
 multiplets  of the  supergravity 
 modes,    in full analogy with what happened  for $(-1)$  term in  the   $\N=4$ $SU(N)$ SYM case 
 \ci{Beccaria:2014xda}.  
 This demonstrates  again     that all long multiplets of 
 massive string modes   do not contribute to  the  conformal anomaly in the
  maximally supersymmetric case.

First,  let us recall  the  KK   spectrum of type IIB supergravity on 
$AdS_{5}\times S^{5}$  \cite{Gunaydin:1984fk,Kim:1985ez}. It is  shown in Table \ref{KK}   where  for  each  KK  level $p$  we list  the   corresponding 
   $SO(2,4)$  and      $SU(4)$ representations  (we use  the same notation as in 
\cite{Beccaria:2014xda}). 
   \begin{table}[t]
\be
\begin{array}{|c|c|c|c}
\hline
 & (\Delta;\, j_{1},j_{2}) & SU(4)  & \\
 \hline
 & (p;\,0,0) & (0,p,0) & \\
 p \ge 1& (p+\frac{1}{2};\,\frac{1}{2},0) & (0,p-1,1)_{c} &\\
 & (p+1;\,1,0) & (0,p-1,0)_{c}  & \times \\
 \hline
& (p+1;\,0,0) & (0,p-2,2)_{c}  & \times \\
 & (p+2;\,0,0) & (0,p-2,0)_{c}  &\\
p\ge 2   & (p+\frac{3}{2};\,\frac{1}{2},0) & (0,p-2,1)_{c} & \\
 & (p+1;\,\frac{1}{2},\frac{1}{2}) & (1,p-2,1)  &  \\
 & (p+\frac{3}{2};\,1,\frac{1}{2}) & (1,p-2,0)_{c} & \\
 & (p+2;\,1,1) & (0,p-2,0)  &\\ 
 &&& \\
 &&& \\
 &&& \\
\hline
\end{array}
\begin{array}{||c|c|c|c|}
\hline
 & (\Delta;\,j_{1},j_{2}) & SU(4)  & \\
 \hline
   & (p+\frac{3}{2};\,\frac{1}{2},0) & (2,p-3,1)_{c} & \\
  & (p+\frac{5}{2};\,\frac{1}{2},0) & (0,p-3,1)_{c}  &\\
p\ge 3  & (p+2;\,\frac{1}{2},\frac{1}{2}) & (1,p-3,1)_{c} &\times  \\
  & (p+2;\,1,0) & (2,p-3,0)_{c}&  \\
  & (p+3;\,1,0) & (0,p-3,0)_{c} & \times \\
  & (p+\frac{5}{2};\,1,\frac{1}{2}) & (1,p-3,0)_{c}&  \\
  \hline
  & (p+2;\,0,0) & (2,p-4,2)  &\\
  & (p+3;\,0,0) & (0,p-4,2)_{c} & \times \\
p\ge 4  & (p+4;\,0,0) & (0,p-4,0)  & \\
  & (p+\frac{5}{2};\,\frac{1}{2},0) & (2,p-4,1)_{c} & \\
  & (p+\frac{7}{2};\,\frac{1}{2},0) & (0,p-4,1)_{c} & \\
  & (p+3;\,\frac{1}{2},\frac{1}{2}) & (1,p-4,1)  & \\
  \hline
  \end{array}
\nonumber
\ee
\caption{\small Multiplet   content of  compactification of type  IIB supergravity on $S^5$. Here  $\Delta$ is   conformal dimension  and  $(a,b,c)$ are Dynkin labels of $SU(4)$.  We use the notation 
 $(a,b,c)_{c} = (a,b,c)+(c,b,a)$ with a corresponding swap of   $(j_{1}, j_{2})$ indices.
A cross means that the corresponding 
 mode originates
 from the 10d   tensor  which is  a complex combination of the NS-NS and R-R 
rank 2  antisymmetric  tensor   fields 
  \cf \cite{Kim:1985ez}.}
\label{KK}
\end{table}
 The   dimension of $SU(4)$  representation $(a,b,c)$  is given by 
\be
\label{12.1}
\dim(a,b,c)=\tfrac{1}{12}(a+1)(b+1)(c+1)(a+b+2)(b+c+2)(a+b+c+3) \ .
\ee
The level $p=1$  corresponds to the doubleton multiplet  which is 
  decoupled from the  physical  spectrum (the corresponding  states  are pure-gauge ones).
  The level $p=2$ is the massless multiplet  of  gauged  $\N=8$ 5d supergravity. 
  The states with $p\ge 3$  form  shortened  massive multiplets with spin $\le 2$. 
  
Applying the  orientifold projection leading  to AdS$_5{\times}  \mathbb{RP}^5$
 involves modding out by the $\mathbb Z_{2}$ subgroup of the $U(1)_{R}$ in the decomposition $SU(4)\supset SU(2)\times SU(2)\times U(1)_{R}$. 
 In addition, the orientifold acts non-trivially on the supergravity fields changing sign 
 of  states originating 
 from the 10d  rank 2 antisymmetric tensor. This means the projection based on the  value of the 
 $U(1)_R $  charge $Q_{R}$ \cite{Fayyazuddin:1998fb}
 \ba\la{1122}
 (j_{1}, j_{2}) \neq (1,0), (0,1): \ \ \ \ Q_{R} &= 0\, (\text{mod}\, 4) \ , \\
 (j_{1}, j_{2}) = (1,0), (0,1): \  \ \ \ Q_{R} &= 2\, (\text{mod}\, 4) \ .
 \ea
For  example, for  the $SU(4)$ representations of the form $(0,p,0)$
 with  $p=0,1,2,3 $ one finds  
\ba 
(0,0,0) &= \bm{1} = (\bm{1},\bm{1})_{0}, \no \\
(0,1,0) &= \bm{6} = \branching{1}{1}{\pm 2}+\branching{2}{2}{0}, \no \\
(0,2,0) &= \bm{20'} = \branching{1}{1}{0}+\branching{1}{1}{\pm 4}+\branching{2}{2}{\pm 2}+\branching{3}{3}{0},\no  \\
(0,3,0) &=\bm{50} = \branching{1}{1}{\pm 2}+\branching{1}{1}{\pm 6}+\branching{2}{2}{0}+\branching{2}{2}{\pm 4}+\branching{3}{3}{\pm 2}+\branching{4}{4}{0}\ . \la{4411}
\ea
Here in the r.h.s. we labelled representations by the dimensions of the two $SU(2)$ representations and indicated  the $R$-charge  ($\pm$ means the  sum over the two  values of the sign).
Note also that  since the  $\Z_{2}$ action on $S^{5}$ giving  $\mathbb{RP}^5$  is free so that  there are 
no  twisted-sector  states. 

To compute the  1-loop  partition function 
 we are   to sum over the  states at each KK level $p$  and then over levels $p$. 
 Let us 
introduce the notation:
\be\no
\dim (a,b,c)|_{q} = \text{sum of dimensions of branched reps with the constraint}\  Q_{R} = q\, (\text{mod}\, 4) \ .
\ee
 We have found the following explicit expressions
\ba
\dim (0,p,0)|_{0} &=\te \frac{1}{16} (-1)^p+\frac{1}{48} (45+80 p+52 p^2+16 p^3+2 p^4), \notag \\
\dim (0,p,1)|_{0} &=\te \dim (1,p,2)|_{0} = 0, \notag \\
\dim (0,p,0)|_{2} &=\te -\frac{1}{16} (-1)^p+\frac{1}{48} (3+32 p+40 p^2+16 p^3+2 p^4), \notag \\
\dim (0,p,2)|_{0} &=\te -\frac{3}{16} (-1)^p+\frac{1}{16} (67+144 p+96 p^2+24 p^3+2 p^4),\notag \\
\dim (1,p,1)|_{0} &=\te -\frac{1}{4} (-1)^p+\frac{1}{12} (87+168 p+100 p^2+24 p^3+2 p^4), \notag \\
\dim (1,p,1)|_{2} &= \te\frac{(-1)^p}{4}+\frac{1}{12} (93+168 p+100 p^2+24 p^3+2 p^4), \notag \\
\dim (0,p,2)|_{2} &=\te \frac{3 (-1)^p}{16}+\frac{1}{16} (93+168 p+100 p^2+24 p^3+2 p^4), \notag \\
\dim (2,p,2)|_{0} &= \te\frac{9 (-1)^p}{16}+\frac{1}{16} (695+1120 p+524 p^2+96 p^3+6 p^4)\ .  \la{1210}
\ea
We  can then compute the total  contribution a$_p$ to the a-anomaly  coefficient 
from  all  states at the   level $p$ using the  expression for the a-coefficient 
  derived in \cite{Beccaria:2014xda}.
For $p\ge 4$ we get 
\be
\la{12.10}  
\text{a}_p = \tfrac{1}{8} (-1)^p p+\tfrac{1}{1080}  (521p-704 p^3+470 p^5-152 p^7+12 p^9)\ .
\ee
Like what happened in  the  $AdS_{5}\times S^{5}$ case \cite{Beccaria:2014xda},  this 
expression  \rf{12.10} happens  to  be valid for all $p \geq 1$, i.e. 
it actually 
 reproduces  also the results for  low  values of $p=1,2,3$, even though 
   the structure of states in these cases 
  is  different  from those  for $p \geq 4$.
   In particular, 
due to the orientifolding and changed periodicity on the sphere, 
the  
 states with $p=1$ are no longer  pure-gauge 
 doubleton ones  and thus  their contribution 
 should  be included  in the  sum  
 \cite{Aharony:1998xz}.
 
As in the \adss  case,   the sum  representing the total 1-loop contribution to the a-anomaly  coefficient 
   is divergent  and  thus requires a 
definition  that should be consistent with underlying symmetries of the 10d theory.\foot{
This regularization issue appears due to the procedure of  first compactifying on 
the  5-space and then regularizing; 
it would be absent if the computation were  done  directly in terms of the 10d determinants 
with a covariant regularization
(see also  a discussion in \cite{Beccaria:2014xda}).} 
One  particular regularization (used in similar context  in 
\cite{Mansfield:2003gs,Ardehali:2013xya}) is to  introduce  a factor $z^{p}$ with $z <1$, do the sum and then 
      drop  all  terms that are  singular (power-divergent)  in the   limit   $z\to 1$.
      This way  we get  
\ba
&\aa_{\rm 1-loop} = \sum_{p=1}^\infty \aa_p  \to   \sum_{p=1}^{\infty}\text{a}_{p}z^{p} \Big|_{z\to 1} = \te \frac{4032}{(z-1)^{10}}+\frac{20160}{(z-1)^9}+\frac{123872}{3 (z-1)^8}+\frac{132608}{3 (z-1)^7}+\frac{233962}{9 (z-1)^6}+\frac{7950}{(z-1)^5}\no\\
\te 
&\ \ \ \ \ \ \ \ \ \ \ 
 \te +\frac{90277}{90 (z-1)^4}+\frac{89}{15 (z-1)^3}-\frac{1003}{360 (z-1)^2}+\frac{49}{360 (z-1)}
-\frac{1}{32}+\mc O(z-1) \ \ \to \ \   -\frac{1}{32}\ . \la{616}
\ea
Keeping only the  finite part   of  the sum we thus 
  reproduce  the  1-loop  term $-\frac{1}{32}$  
  in the  conformal a-anomaly in \rf{41}.\foot{To  compare,  in  the $S^5$ compactification  case,  
  using the same regularization 
 one finds  that  $\sum_{p=1}^{\infty}\text{a}_{p}=0$   \cite{Beccaria:2014xda}.  This implies that keeping all KK modes  one gets 
  $\text{a} = \frac{1}{4}N^2$  (with no 1-loop shift)  which is the  result for the conformal anomaly 
  of the $\N=4$ SYM with   $U(N)$ gauge group.
  In the \adss   case dual to $SU(N)$ SYM theory, 
   where  the $U(1)$  multiplet describing the D3-brane  center-of-mass
  degrees of freedom should decouple, the $p=1$  (doubleton)  contribution to the  sum  should  
  not to be included and thus  1-loop   correction to a  should be given by 
  $\sum_{p=2}^{\infty}\text{a}_{p}=-1$. Once again, 
  there is  no similar  decoupling of the $p=1$   level in the  present  orientifold case, i.e. 
  the sum in \rf{616}  starts from  $p=1$. }

The same result is found using an alternative regularization prescription \cite{Beccaria:2014xda}   based on  introducing  an upper  cutoff $P$ in the sum over  $p$ and dropping  all 
divergent terms that are   polynomial  in $P\to \infty$.
Then   the sum of the 
second    $\tfrac{1}{1080}  (521p-704 p^3+470 p^5-152 p^7+12 p^9)
$    term in (\ref{12.10}) gives  a  vanishing contribution.  The remaining 
  sum of the  sign-alternating    
  first term  $ \tfrac{1}{8} (-1)^p\,  p$  in \rf{12.10}  is finite and  
   is  readily   computed    using either
  an  analytic regularization  
\be\la{213}
\alpha \to 1: \ \ \ \ \ 
\tfrac{1}{8}\sum_{p=1}^{\infty}(-1)^{p}\, p^{\alpha} = \tfrac{1}{8}(2^{\alpha+1}-1)\, \zeta(-\alpha) = -\tfrac{1}{32}+\mc O(\alpha-1)\ , 
\ee
or an   exponential cutoff 
\be\la{214}
\eps \to 0: \ \ \ \ \tfrac{1}{8}\sum_{p=1}^{\infty}(-1)^{p}p \, e^{-p\eps}= -\tfrac{e^{\eps}}{8\,(e^{\eps}+1)^{2}} = -\tfrac{1}{32}+\mc O(\eps) \ , 
\ee
in agreement with \rf{616}. 

\subsection{Free energy    in $\N=2$     theory dual to   IIB string on 
AdS$_5{\times}  S^5/\Z_{2,\ori}$ \la{S6-2}}

Let us now attempt  to  give a dual  string theory understanding of 
 the  coefficient of the leading $\log \l$  term in the localization  expression \rf{11} for the  free energy  
of the $\N=2$  $Sp(2N)$   FA theory, i.e. 
\be \la{31}
F= - ( N^2 + N + \tfrac{3}{16}) \log \l + \dots    \ . \ee
We shall   focus on the order $O(N^2)$ and $O(N)$ terms that should come from the sphere and disc/crosscup topologies. The   computation of the remaining   $\tfrac{3}{16}\log \l $ term (that should  come from  the closed-string 1-loop, i.e.  torus  contribution) appears
 to be more challenging  than in the  above   maximally   supersymmetric 
 $\N=4$ SYM case  
 and will not be attempted here.

The main idea is that  as in the $\N=4$ SYM case  \ci{Russo:2012ay}
  the $\log \l$  term in free energy should  be associated   with the regularized 
  expression for  the AdS$_5$ volume factor.
    The order $N^2\log \l $  term 
    originates from the type IIB   supergravity compactified on $S'^5=  S^5/\Z_{2,\ori}$, while 
    the order $N$ term   should come   from the disc/crosscup  contributions that  may be interpreted as the action of D7+O7-branes     wrapped on AdS$_5{\times} S^3$  where $S^3$ is the fixed-point locus 
     of  the orbifold action $\Z_{2,\ori}$.\foot{Here, compared to the $\N=4$ $Sp(2N)$  SYM case discussed above 
       there will  be no shift of $N$ by $\four$    so the order $N$ term in $F$  or in conformal anomaly  will have a different interpretation.}

Let us first  mention that  to reproduce the conformal   anomaly c-coefficient  following \ci{Liu:1998bu, Henningson:1998gx} one   may   consider  the sum of the bulk  10d   supergravity action 
and the  D7+O7 action,  
$S\sim \gs^{-2} \int d^{10} x \sqrt g (R+...) +\gs^{-1}  \int d^8 x \sqrt g   (RR + ...)$ 
assuming that  the 5d  metric  asymptotes  to a general  4d metric at the boundary. 
The on-shell   value of the action then  contains the term 
  $ ( N^2 +  q_1  N) \int d^4 x \sqrt g  C_{mnkl}^2  \log \Lambda $ 
where $C$  is the  Weyl tensor of the 4d   boundary metric  and $\Lambda$ is  
an IR cutoff.
 One finds  \ci{Blau:1999vz} 
that the   order $N$   term coming from the $ \int d^8 x \sqrt g  RR $  action 
 is precisely the one  consistent with the  value of c-a$= \four N$  
 of the $\N=2$  $Sp(2N)$   FA  theory  in \rf{42}.\foot{To  capture the value of the  coefficient  $\cc-\aa$ it is  sufficient to assume that $R_{mn}=0$
for the boundary metric is Ricci-flat  as was effectively  done 
  in \ci{Blau:1999vz}.
  The value of the a-anomaly coefficient was not  reproduced in \ci{Blau:1999vz}
  as only the $R^2_{mnkl}$ term was included in the 8d action.}

To  determine   the a-anomaly 
one may  chose  the round 4-sphere   metric at the boundary (so that  the c-anomaly term proportional to the Weyl tensor squared  will be vanishing).  Then  the on-shell value of the above action  will scale as 
as a  factor of  volume of AdS$_5$ \rf{111}, i.e.  $(N^2 + q_2 N) \int d^5x \sqrt g \sim (N^2 + q_2 N) \log \Lambda_{\rm IR} $. To match the a-anomaly  coefficient  in \rf{42}   one should find that 
$N^2 + q_2 N = N^2 + N$.   

The  computation of the $\log \l$ term in the free  energy $F$ on the dual string theory side is essentially equivalent  to the computation of the a-anomaly term  if we assume, following 
\ci{Russo:2012ay},  that the $\log \l$ originates from a particular regularization of the AdS$_5$ volume factor as in \rf{111}.
This effectively explains, on the dual string theory side, why the $N^2 + N$ terms in the a-anomaly 
in \rf{42} happens to be the same as in the $\log \l$ term in the free energy $F$ in \rf{31}.

In the case of the $Sp(2N)$ FA  model  dual to type IIB string on the orientifold
 AdS$_5 \times S^{5}/\mathbb Z_{2, \rm ori}$ we  start with $N_{\rm cover}=2N$   D3-branes   so that (cf. \rf{1xx}) 
\be \la{33} 
L^{4} =  \tfrac{\sqrt\pi\, }{2}\,\kappa_{10} \, \frac{N_{\rm cover}}{\Vol(S^{5})}=  8\pi g_{s}N\alpha'^{2}
\ .  \ee 
As a result, as in the $\N=4$  $Sp(2N)$ SYM case  
\be\la{303} S_{10} =     \tfrac{2}{ \pi^2}  N^2 \text{Vol}({\rm AdS}_{5}) =    2 N^{2}\,\log(\L_{\rm IR}\rr) \ , \qquad \ \ \   \aa= \half  N^2 + ... \ , \ee
in agreement with \rf{42}.

To try to  find the  subleading order $N$ term in  free energy 
we shall as in   \ci{Blau:1999vz} 
  add to the bulk 10d supergravity action (2-sphere  contribution) the  8d integral  of the 
   effective action $S_7$ of a combination of 8 D7-branes  and an orientifold plane (disc+crosscup contribution).
 The presence of the orientifold plane cancels  the tension part of $S_7$ (and thus there is no  dilaton tadpole, implying stability consistent with   supersymmetry). As a result,    
  $S_7$ starts with terms  quadratic in the curvature.\foot{ 
In general, for  $n$    D$_p$-branes   and an orientifold $p$-plane    we get 
a combination $n \, {\rm D}_p - 2^{p-4} \, {\rm O}_{p}$  of DBI+WZ   actions 
while for the  curvature-squared terms  we get 
$n \, {\rm D}_p +  2^{p-5} \, {\rm O}_{p}$. In  the present  $p=7$ 
case we need $n=8$  to cancel the tadpole term  and thus 
 the $R^{2}$  term   enters  with  the overall  coefficient 
$8 + 4=12$.
}
Then  according to \cite{Bachas:1999um} (see also  \ci{Fotopoulos:2001pt,Schnitzer:2002rt,Junghans:2014zla,Weissenbacher:2020cyf}) 
\be\la{34}
S_{7} = k_7 
\,\int d^{8}x\sqrt g \   \LL_{8}\ , \qquad 
k_7 = 12  \frac{(2\pi\alpha')^{2}}{6\times 32g_{s}}\,\mu_{7} \ , \qquad 
\mu_{7} = \frac{1}{2\,(2\pi)^{7}\alpha'^{4}},
\ee
where $\mu_7$ is the D7-brane tension\foot{In the case of an  orientifold 
 the value of the D7 brane tension  is reduced by $\ha$ (see, e.g.,  \cite{Aharony:1999rz}).}
   and 
\be \la{35} \LL_8= \LL_{R^2} + \LL_{F_5} \ , \qquad 
 \LL_{R^2} = (R_{\rm T})_{mnkl}^{2}-2\,(R_{\rm T})_{mn}^{2}+2\,{\bar R}_{ab}^{2} \ .
\ee 
Here 
the tensor $(R_{\rm T})_{mnkl}$ is the pull-back of Riemann tensor on the brane tangent bundle, $(R_{\rm T})_{mn}$ is its Ricci tensor contraction
involving only tangent indices, and ${\bar R}_{ab}$ is the tangent space  contraction of the Riemann tensor with normal bundle indices  $(a,b)$ (see \cite{Bachas:1999um, Junghans:2014zla,Weissenbacher:2020cyf} for details).  $\LL_{F_5}$  in \rf{35} stands
for RR 5-form  dependent terms that were not  determined  in \cite{Bachas:1999um}.\foot{These terms could    be,  in principle,   extracted from the 
six-point open string  amplitudes  on the disk and  crosscup. 
The knowledge of 
$\LL_{F_5}$  is not needed to compute the c-a anomaly  coefficient not 
  sensitive to the  Ricci tensor 
 of the boundary  metric  -- as was already mentioned above, 
  the leading  order $N$   term in  $\rm c-a$= $\four N - {1\ov 12}$ (cf. \rf{42}) 
 was  reproduced in  \ci{Blau:1999vz}   just from  the  knowledge of the 
 $ R^2_{mnkl}$ term in $ \LL_{R^2}$.}
 We ignore other normal   bundle  contributions that are  vanishing  
  in the present case. 
 
 Let us note   that \ci{Blau:1999vz}  considered only the first $R^2_{mnkl}$ term 
 in $\LL_{R^2}$   while  ref. \ci{Guralnik:2004ve} discussed  the $F^2_{mn}\LL_{R^2}$ term in the 
 D7-brane action in the context of  investigation  of the  holographic dual of the Higgs branch 
 of the $\N=2$ $Sp(2N)$  FA theory.\foot{In this case one may view D3-branes as 
 instantons in 8d theory  describing D7-branes. It was suggested in \ci{Guralnik:2004ve} 
 that  the condition  of existence of uncorrected   gauge-theory instanton   solution  imposes constraints on the structure of possible $\a'$-corrections of the type  $F^2_{mn} {\cal F} ( R, F_5)$. The $F_5$ dependent terms   here need not be a priori the same as appearing in $\LL_8$  in \rf{35}   so the vanishing of $F^2_{mn} {\cal F} ( R, F_5)$  on  AdS$_5 \times S^3$    background 
 need not imply the same for \rf{34}.}

 In the present case D7+O7 branes   are  
wrapped on AdS$_5{\times} S^3$  where $S^3$ is the fixed-point locus  of $\Z_{2,\ori}$.
Normalizing the metric and curvature to unit scale, i.e.  extracting a factor of AdS  radius $L$ 
 in \rf{33}   we get  for the coefficient in \rf{34}
\be \la{36} 
k_7\  \to \  k'_7= L^4\,   k_7 = 12\  \frac{( 2\pi \alpha')^2}{6\times 32  g_s}\ \frac{1}{2\,(2 \pi)^7 \alpha'^{4}}\  (  8 \pi g_s N  \alpha'^2)
 = \frac{N}{128\pi^{4}}.
\ee 
Ignoring $\LL_{F_5} $  in \rf{35} and using that the curvature of AdS$_5{\times} S^3$  is  homogeneous 
we should   have 
\be \la{37}
\int d^{8}x\sqrt g \   \LL_{R^2}= \Vol({\rm AdS_5})\,  \Vol(S^3) \,   \K_{{R^2}} \ .
\ee
 Combining the $S_7$ term \rf{34} with the  bulk supergravity term \rf{303}  we get  ($\Vol(S^3)= 2 \pi^3$)
\be
\la{38}
S =S_{10} + S_8=  \tfrac{2}{\pi^2} \Vol(AdS_{5})\, 
\big(N^{2}-\tfrac{1}{128} \K_{{R^2}}  N \big).
\ee
To compute the coefficient $\K_{{R^2}} $ in \rf{37} 
we need to account  for the curvature of AdS$_5{\times} S^3$.
For unit-radius  curvature of $AdS_{5}\times S^{3}$  we have 
\be\la{306}
(R_{\rm T})_{mnkl} = \begin{cases}
\mp(g_{mk}g_{nl}-g_{ml}g_{nk}), &  \text{all indices in }\, AdS_{5}\ \text{or in }\ S^{3} \\
0, & \text{mixed indices}.
\end{cases}
\ee
In $d$ dimensions  $(R_{\rm T})_{mn} = \pm(d-1)g_{mn}$ and thus
\be\la{307}
(R_{\rm T})_{mnkl}^{2} -2\,(R_{\rm T})_{mn}^{2} = 2d(d-1)-2 d(d-1)^{2} = -2d(d-1)(d-2) = 
\begin{cases}
-120, & AdS_{5}, \\
-12, & S^{3}.
\end{cases} 
\ee
Finally,  $\bar R_{ab} = g^{mn}R_{mabn} = -3g_{ab}$ with $m,n=1,\dots,8$ (tangent bundle) and 
$a,b=9,10$ (normal bundle), so that  $ 2 \bar R_{ab} ^2 = 2\times 3^{2}\times 2 = 36 $.
The total   value of the coefficient  $\K_{{R^2}} $   in \rf{37},\rf{35}   is then\footnote{This  value 
of the $R^2$  coefficient  is in agreement with the one   found in  \cite{Guralnik:2004ve}.}
  \be \la{308}
\K_{{R^2}}   = -120 - 12+  36  = -96\ . \ee
As a result, eq. \rf{38}   becomes 
\be\la{39} 
S =  \tfrac{2}{\pi^2}\Vol(AdS_{5})\,  \big(N^{2}+\tfrac{3}{4}N\big) \ . 
\ee
This   differs   from the expected 
 $N^{2}+N = N^{2}+\frac{3}{4}N +\frac{1}{4}N$, \ie we  are missing 
  an extra $+\frac{1}{4}$
 contribution to  $\K_{{R^2}} $. 
 
This  discrepancy with the expected result for the $N^2 + N$ terms  in the 
 free energy \rf{31} and the conformal a-anomaly \rf{42}
\be\la{399} 
F =  \tfrac{2}{\pi^2}\Vol(AdS_{5})\,  \big(N^{2}+N\big)=   2 (N^2 + N) \log ( \L_{\rm IR} 
 \rr) \to - (N^2 + N) \log ( \l    \, \rr^{-2})\ 
\ee
may be attributed to the fact that we did not take into 
account the $F_5$-dependent terms in \rf{35}.
 In the present case of AdS$_5 \times S'^5$   background
 possible  $F_{5}$-dependent  terms 
may effectively  contribute similarly to the Ricci-squared terms.
More precisely,  in  the derivation
of $\LL_{R^2}$ term  in  \cite{Bachas:1999um}  it was assumed  that the bulk space curvature is 
 Ricci flat,   $R_{MN}=0$, and $F_5$-dependent terms were ignored. The two  types of terms are actually related by the 10d supergravity equations
         $
R_{MN} = \frac{1}{4\times 4!} F_{MPQRS}F\indices{_{N}^{PQRS}}, 
\ \   R_{MN}^{2} \sim (F_{MPQRS}F\indices{_{N}^{PQRS}})^2$. 
Thus  the missing terms   may be parametrised as $R_{MN}^{2}$, implying 
  the following possible  correction on AdS$_5 \times S^3$  background 
\be\la{400}
\LL_{F_5}= {\rm k}_F\, R_{MN}^{2} = 160\,{\rm k}_F\ .
\ee
Then  \rf{39}   becomes the  $N^{2}+N$  combination in \rf{399}
if   ${\rm k}_F=\frac{1}{5}$.
Proving that this is actually the case remains an open problem. 

\subsubsection*{Acknowledgments}

\small {We are thank to  S. Giombi, A. Hanany,  J. Minahan  and A. Vainshtein  for  useful   
discussions. We are also  grateful    S. Chester and  G. Dunne  for   helpful comments on the draft.
 MB was supported by the INFN grant GSS (Gauge Theories, Strings and Supergravity).
The work of GK was supported by the French National Agency for Research grant ANR-17-CE31-0001-01.
AAT was supported by the STFC grant ST/T000791/1. 
Part of this   work was done while AAT was a participant of the ``Integrability in String, Field, and Condensed Matter Theory'' at   KITP in Santa  Barbara    supported in part by the NSF under Grant No. NSF PHY-1748958.
}
 
 \newpage
 \tiny{
\bibliography{BKT-Biblio}
\bibliographystyle{JHEP}
}
\end{document}